\documentclass[reprint,superscriptaddress,nofootinbib,amsmath,amssymb,aps,longbibliography]{revtex4-2}

\usepackage{graphicx}
\usepackage{dcolumn}
\usepackage{bm}

\usepackage{bbm}
\usepackage{bbold} 
\usepackage{hyperref}
\hypersetup{
    colorlinks=true,
    linkcolor=black,
    citecolor=blue,
} 
\usepackage{tikz}

\newcommand{\bra}[1]{\langle#1\vert}
\newcommand{\ket}[1]{\vert#1\rangle}
\newcommand{\bracket}[2]{\langle#1\vert#2\rangle}
\newcommand{\bracketmatrix}[3]{\langle#1\vert#2\vert#3\rangle}

\newcommand{\xup}{{\uparrow}}
\newcommand{\xdown}{{\downarrow}}

\DeclareMathOperator{\Tr}{Tr}

\makeatletter
\newcommand{\phantomlabel}[2]{
    \protected@write\@auxout{}{
        \string\newlabel{#2}{
            {\@currentlabel#1}{\thepage}
            {\@currentlabel#1}{#2}{}
        }
    }
    \hypertarget{#2}{}
}
\makeatother

\definecolor{mygreen}{rgb}{0.0,0.55,0.3}

\definecolor{calumblue}{RGB}{10, 168, 199}

\definecolor{calumpurple}{RGB}{100, 100, 230}

\definecolor{calumred}{RGB}{200, 30, 30}

\usepackage{soul}
\setstcolor{calumred}

\usepackage{nicematrix}

\newcommand{\beq}{\begin{equation}}
\newcommand{\eeq}{\end{equation}}

\begin{document}

\title{Unravelling Metastable Markovian Open Quantum Systems}

\author{Calum A. Brown}
\affiliation{Department of Applied Mathematics and Theoretical Physics, University of Cambridge, Wilberforce Road, Cambridge CB3 0WA, United Kingdom}
\author{Katarzyna Macieszczak}
\affiliation{Department of Physics, University of Warwick, Coventry CV4 7AL, United Kingdom}
\author{Robert L. Jack}
\affiliation{Department of Applied Mathematics and Theoretical Physics, University of Cambridge, Wilberforce Road, Cambridge CB3 0WA, United Kingdom}
\affiliation{Yusuf Hamied Department of Chemistry, University of Cambridge, Lensfield Road, Cambridge CB2 1EW, United Kingdom}

\begin{abstract}
We analyse the dynamics of metastable Markovian open quantum systems by unravelling their average dynamics into stochastic trajectories. We use quantum reset processes as examples to illustrate metastable phenomenology, including a simple three-state model whose metastability is of classical type, and a two-qubit model that features a metastable decoherence free subspace. In the three-state model, the trajectories exhibit classical metastable phenomenology: fast relaxation into distinct phases and slow transitions between them.  This extends the existing correspondence between classical and quantum metastability.  It enables the computation of committors for the quantum phases, and the mechanisms of rare transitions between them.  For the two-qubit model, the decoherence-free subspace appears in the unravelled trajectories as a slow manifold on which the quantum state has a continuous slow evolution.  This provides a classical (non-metastable) analogue of this quantum effect.  We discuss the general implications of these results, and we highlight the useful role of quantum reset processes for analysis of quantum trajectories in metastable systems.
\end{abstract}

\maketitle

\hypersetup{
    colorlinks=true,
    linkcolor=magenta,
    citecolor=blue,
}


\section{INTRODUCTION}

Isolated quantum systems evolve unitarily, according to the Schr\"odinger equation.  However, practical quantum systems are never isolated -- their unavoidable coupling to the environment leads to effects such as dissipation and decoherence~\cite{Breuer_and_Petruccione,quantum_noise}. Indeed, recent experimental advances have demonstrated a rich phenomenology in open quantum systems,  such as ultracold atoms \cite{song_22}, optomechanical systems \cite{Cavity_optomechanics_review,Gil-Santos_2017}, superconducting circuits \cite{Houck_2012,Fitzpatrick_2017} and Rydberg atoms in optical lattices \cite{Muller_2012,Bernien_2017}.   Understanding and modelling these systems is essential for applications of quantum technologies.

Metastability in open quantum systems occurs when their dynamical relaxation features a separation of timescales \cite{Macieszczak2016,Minganti2018,classical_metastability_in_quantum,Macieszczak2021operational}.  This offers a route to realisation of long-lived quantum coherences \cite{Letscher2017,Souto2017,Jessop2020,Matern2023}, which are crucial for quantum technologies such as computation \cite{Senko_2020,Allcock_2021}, {which require quantum memories \cite{Labay-Mora23,experimental_DFS_2001}}.
Understanding emergent slow time scales is also important for more fundamental questions that arise in {open quantum systems, including quantum phenomena associated with non-equilibrium phase transitions \cite{Diehl_2010,Cabot_2021,Landa20,Gambetta2019}.  These include metastable decoherence free subspaces (DFSs)~\cite{Macieszczak2016} and glassy phenomenology~\cite{facilitated_spin_models,meta_east}.}

This work focuses on Markovian metastable open quantum systems, which exhibit a rich phenomenology, while remaining theoretically tractable.  Recent work~\cite{olmos2014,Thingna2016,Gambetta2019,meta_ising}
has analysed such systems through the quantum master equation (QME) which describes deterministic non-unitary evolution of the system's density matrix~\cite{Lindblad_1976,Gorini1976CompletelyPD}.  This allows characterisation of metastable phenomena such as slow transient relaxation to the steady state, and slow decay of steady-state autocorrelation functions~\cite{Macieszczak2016}.  
However, the resulting theory of quantum metastability is distinct from its classical counterpart: the quantum theory gives access to expectation values but the classical theory  also predicts the behaviour of stochastic trajectories.
The behaviour of trajectories in the quantum setting requires information beyond the QME, including time records of environmental measurements~\cite{Observation_of_quantum_jumps,bergquist1986,exp_weak_1,exp_weak2,Blok2014,Lange2014} and the unravelled quantum state, via a quantum trajectory formalism~\cite{Belavkin1990,Dalibard1992,Gardiner1992,Carmichael1993AnOS}.
Analysis of these trajectories reveals phenomena beyond the reach of the QME, including full-counting statistics of photon emissions~\cite{thermodynamics_quantum_jump_trajectories,znidaric_2014,Carollo2019,Liu2022}, measurement-induced phase transitions \cite{Li2018,Chan2019,Skinner2019,Turkeshi2021,passarelli2023}, geometric phase transitions \cite{Cho2019,gebhart2020}, the quantum Zeno effect \cite{Snizhko20,Biella2021,Walls_2022}, quantum steering \cite{active2021}  and quantum thermodynamics \cite{Hekking2013,Manzano2022}.

This work analyses
several metastable quantum systems in the trajectory formalism.
This extends previous results based on the QME~\cite{Macieszczak2016,classical_metastability_in_quantum}, and time records~\cite{classical_metastability_in_quantum}, and strengthens the connections to classical theories of metastability.  For example, metastable classical systems relax quickly into distinct phases, after which they make slow transitions between them; one may also identify basins of attractions of the phases, via the committor~\cite{PetersBook,E2010}.
Our analysis of quantum trajectories yields corresponding committors for metastable quantum phases, as well as the mechanism of transitions between them. This allows characterisation of fluctuations within these systems' nonequilibrium steady states, beyond the QME.

We demonstrate these results by  
analysing several different systems, which highlight general features of metastability, as well as providing their own specific insights.  We mostly focus on quantum reset models~\cite{Carollo2019,Carollo2021}, which are particularly tractable in the trajectory formalism.  These models have the property that each quantum jump operator has a unique destination state, which are termed reset states.  While this may seem like a significant restriction, such models are known to be rich enough to support classical metastable behaviour (in the sense of~\cite{Macieszczak2016,classical_metastability_in_quantum}).  We show here that they also exhibit intrinsically quantum metastability, specifically DFSs.  

\begin{figure}
    \centering
    \includegraphics[width=0.48\textwidth]{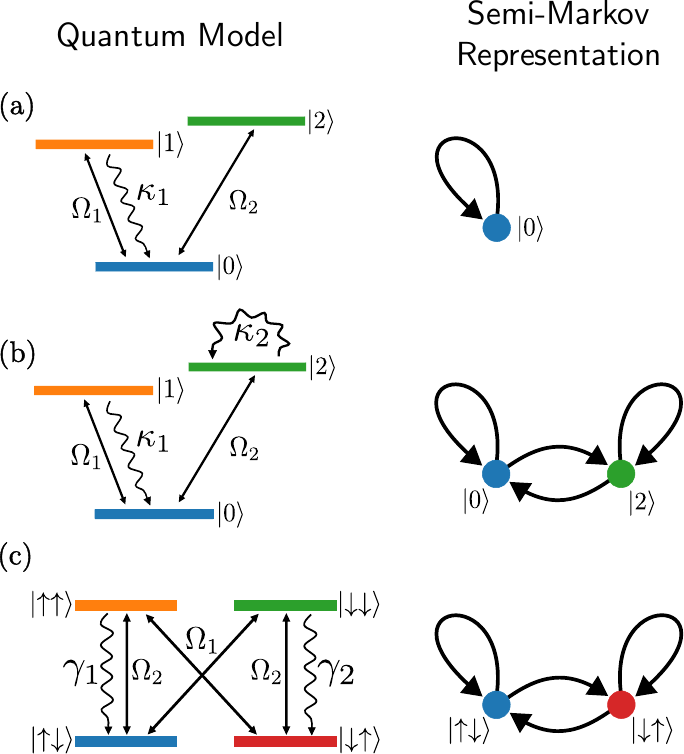}
    \caption{Example models studied in this work.  (a)~Three-state model with a single jump operator, see Sec.~\ref{sec:three_state_intro}. (b)~Three-state model with two jumps, see Sec.~\ref{sec:3_state_2_jump}. (c)~Two-qubit model featuring a metastable DFS, see  Sec.~\ref{sec:two_qubit_model}.}
    \label{fig:model_diagrams}
    \phantomlabel{(a)}{fig:3_state:sys_diagrams}
    \phantomlabel{(b)}{fig:two_jump:sys_diagrams}
    \phantomlabel{(c)}{fig:dfs_reset:sys_diagrams}
    \phantomlabel{(a,b)}{fig:3_state_diagrams_both_models}
\end{figure}

Quantum reset models are convenient for trajectory analyses because their unravelled dynamics can be restricted to a relatively small set of quantum states, avoiding the requirement to follow the complicated random time evolution of a  wavefunction or density matrix.  In fact, many aspects of these systems can be obtained from a mapping to classical semi-Markov processes~\cite{Carollo2019,Carollo2021}: this means that the rate of quantum jumps at time $t$ only depends on the destination of the last jump, and the time elapsed since that jump. In this work, we discuss several examples within this class, which are depicted in Fig.~\ref{fig:model_diagrams}. The left column shows the quantum models, with solid arrows indicating unitary (Hamiltonian) evolution, and the wavy arrows quantum jumps.  The right column shows the semi-Markov representation, whose internal states correspond to the reset states of the jumps; the arrows indicate the jumps, which may reset the state to the previous destination, or jump to a new one.

The three models are illustrative of several aspects of quantum metastability.  The models in Fig.~\ref{fig:3_state_diagrams_both_models} each have a three-dimensional Hilbert space, but they differ in the number of quantum jump operators, and thus in the number of reset states in the semi-Markov representation.  For suitable parameters, both models exhibit classical metastability in the sense of~\cite{Macieszczak2016,classical_metastability_in_quantum}.  This work additionally shows that their quantum trajectories support the full metastable phenomenology expected for classical systems, such as fast relaxation into the metastable phases, and slow transitions between them.  We also identify the mechanisms for transitions between the metastable phases, and we analyse the committors for the phases.  This material is the subject of Sec.~\ref{sec:three_state_models}.
%
Based on these observations, we then present in Sec.~\ref{sec:metastable_unravelled} several general results for the committor, and its connections with the QME.

In contrast to these three-state models, the two-qubit (four-state) system of Fig.~\ref{fig:dfs_reset:sys_diagrams} supports a non-classical form of metastability -- a metastable DFS -- which manifests in quantum trajectories as a manifold on which the quantum state evolves in a slow but continuous fashion, although the number of reset states is finite.  This differs qualitatively from classical metastability in the sense of~\cite{classical_metastability_in_quantum}, which features rare transitions between discrete metastable phases.  This model is discussed in Sec.~\ref{sec:two_qubit_model}.

Throughout this work, we use these example systems to identify and explain generic features of trajectories of metastable open quantum systems.  While previous work has focussed on the QME evolution and on experimental time records~\cite{classical_metastability_in_quantum,meta_east}, our focus on quantum trajectories gives a more direct connection to metastability in classical systems, for which the focus on trajectories is natural, as in transition state theory and transition path theory~\cite{PetersBook,E2010}. 
While quantum trajectories are less intuitive than their classical counterparts, our examples of quantum reset processes result in simple and physically-informative descriptions. 

Our concluding Sec.~\ref{sec:outlook} summarises our main insights and surveys open directions.  
For quantum metastability with discrete phases, we discuss a detailed correspondence with metastability of classical systems, including the committor, the intermittent fluctuations of the quantum state, and the mechanisms of transitions between metastable phases.  For intrinsically quantum metastable phenomena like DFS, the results establish a different kind of correspondence with classical stochastic processes, in terms of slow continuous relaxation that would not be interpreted as metastability in the classical setting, but rather as slow relaxation within a continuous manifold.  We also explain that while our examples have been taken from quantum reset processes, many of these conclusions are generic for metastable open quantum systems.

\section{Dynamics of Markovian open quantum systems}\label{sec:general_theory}

This Section summarises theoretical background for metastable open quantum systems, and the concept of the committor from classical metastable systems.

\subsection{Quantum master equation}\label{sec:markovian_oqs_dynamics}

The quantum master equation (QME) \cite{Lindblad_1976,Gorini1976CompletelyPD,davies1974} is a generic description of the dynamics of the density matrix $\rho$ of a Markovian open quantum system.  It takes the form
\begin{align}\label{Lindblad_equation}
    \dot{\rho} & = \mathcal{L}(\rho) \; ,
    \nonumber\\
    \mathcal{L}(\rho) & = -i[H,\rho] + \sum_{k=1}^M \left(J_k\rho J_k^\dagger - \frac{J_k^\dagger J_k \rho + \rho J_k^\dagger J_k}{2} \right), 
\end{align}
where $H$ is the system Hamiltonian and the $J_k$ are jump operators which describe the environmental interaction.  We consider systems defined on finite-dimensional Hilbert spaces $\cal H$.

The linear operator $\cal L$ is called the {quantum Liouvillian}. Its  eigenvalues have non-positive real parts, they are denoted $\lambda_j$ for $j=1,2,\dots,{\rm dim}({\cal H})^2$.  We order them according to their real parts: ${\rm Re}(\lambda_j) \leq {\rm Re}(\lambda_{j-1}) \leq \cdots \leq \lambda_1$.  The corresponding eigenmatrices of ${\cal L}$ are denoted by $R_j$ and those of ${\cal L}^\dag$ are denoted by $L_k$.  These satisfy $\Tr\left[ L_i R_j\right] = \delta_{ij}$.\footnote{%
We assume for simplicity that ${\cal L}$ is diagonalisable, ignoring the possibility of Jordan blocks.}
We assume that the open quantum system has a unique steady state, so there is a non-degenerate zero eigenvalue $\lambda_1=0$ and the density matrix for this steady state is $\rho_\textsc{ss}=R_1$.

\subsection{Metastability in the QME}\label{sec:metastability}

\newcommand{\taumet}{\tau_\textsc{s}}
\newcommand{\tauf}{\tau_\textsc{f}}
\newcommand{\commet}{C^*}

In the QME description, a quantum system is metastable if its spectrum has a gap: the operator ${\cal L}$ has $m$ ``slow'' eigenvalues which separate from the rest of the spectrum.  There are slow and fast time scales associated with metastability:
\beq
\taumet = \frac{-1}{{\rm Re}(\lambda_m)}, \qquad \tauf= \frac{-1}{{\rm Re}(\lambda_{m+1})}
\eeq
where the gap in the spectrum means that $\taumet \gg \tauf $.

It will be convenient in the following to assume that metastability is controlled by a small parameter $\epsilon$: specifically, we assume that $\lim_{\epsilon\to0} {\rm Re}(\lambda_m)=0$ so that $\taumet$ diverges in this limit, but $\tauf$ is finite.
Hence, after an initial fast relaxation, 
the dynamics  of $\rho(t)$ is controlled by the slow eigenvalues and eigenmatrices of ${\cal L}$~\cite{Macieszczak2016}, such that for $t\gg \tauf$
\begin{equation}\label{MS_approx}
    \rho(t) \approx \rho_\textsc{ss} + \sum_{k=2}^m \alpha_k e^{t\lambda_k}R_k
\end{equation}
where $\alpha_k = \Tr[L_k\rho(0)]$.
This characterisation of the slow dynamics describes the \emph{metastable time regime} $\tauf\ll t\ll \taumet$,\footnote{Mathematically, this regime is accessed by taking first $\epsilon\to0$ and then $t\to\infty$.} as well as the slow relaxation to $\rho_{\textsc{ss}}$ that occurs at long times $t\gg \taumet$ \cite{Baumgartner_2008}.

Consider two-phase metastability, $m=2$, which has many similarities with metastablility in simple classical systems. The emerging picture is that the system has two distinct metastable phases which are described by matrices $\rho_A,\rho_B$ which have unit trace.  These can be computed as
\begin{equation}
    \rho_A = \rho_{\textsc{ss}}+\alpha_2^{\text{max}}R_2, \quad \rho_B = \rho_{\textsc{ss}}+\alpha_2^{\text{min}}R_2,
    \label{equ:rhoAB}
\end{equation}
where $\alpha_2^{\text{max}}$, $\alpha_2^{\text{min}}$ are the extremal eigenvalues of $L_2$ \cite{classical_metastability_in_quantum}.  For small $\epsilon$, the matrices $\rho_A,\rho_B$ are called are called extremal metastable states (EMS), they are \emph{almost} density matrices, as explained in~\cite{Macieszczak2016,classical_metastability_in_quantum}.

The emerging physical picture is that $\rho(t)$ relaxes quickly into a linear combination of $\rho_A,\rho_B$: during the metastable time regime $\tauf\ll t\ll \taumet$ the system appears stationary because \eqref{MS_approx} reduces to
\beq
    \rho(t) \approx p_A \rho_A + p_B \rho_B
    \label{equ:rhoAB-meta}
\eeq
where $p_{A,B}={\rm Tr}[P_{A,B}\rho(0)]$, with 
\beq
P_A =  \frac{L_2-\alpha_2^\text{min}\mathbb{1}}{\alpha_2^\text{max} - \alpha_2^\text{min}}
, \qquad
P_B =  \frac{-L_2+\alpha_2^\text{max}\mathbb{1}}{\alpha_2^\text{max} - \alpha_2^\text{min}} \; .
\label{equ:proj}
\eeq
These operators have non-negative eigenvalues and $P_A+P_B=\mathbb{1}$, ensuring that $p_A,p_B$ are probabilities.\footnote{%
$P_A$ and $P_B$ form a positive operator-valued measure (POVM).}  
This means in particular that a
system initialised in a state $\rho(0)$ with ${\rm Tr}[P_A\rho(0)]=1$ will relax quickly to $\rho_A$ (on timescale $\tauf$).
The approximate equality in \eqref{equ:rhoAB-meta} appears because we have neglected terms that vanish as either $\tauf/t\to0$ or $t/\taumet\to0$.

{The situation described here for $m=2$ corresponds to classical metastability in the sense of~\cite{Macieszczak2016,classical_metastability_in_quantum}: within the metastable time regime, $\rho(t)$ can be expressed as a linear combination of density matrices corresponding to distinct phases, with real coefficients that correspond to probabilities, as in~\eqref{equ:rhoAB-meta}.  More complex forms of quantum metastability  -- such as DFS -- can occur when $m>2$, these involve long-lived coherences between the phases,} see Sec.~\ref{sec:two_qubit_model}.

\subsection{Trajectories of quantum reset processes and their mapping to semi-Markov processes}
\label{sec:semi-Markov}

The QME describes the evolution of the quantum system, while the environment has been integrated out completely.  
Quantum trajectory theory \cite{Molmer93,Carmichael1993AnOS,Breuer_and_Petruccione,Wiseman_Milburn,Daley_2014} describes systems subject to continuous monitoring, including the joint statistics of measurements in the system and the environment.
For example, consider a driven system in which experiments yield a time record of photon emissions, as observed in  quantum optics experiments \cite{Observation_of_quantum_jumps,nagourney1986,bergquist1986}, superconducting qubits \cite{exp_weak2,Weber_2016} and quantum dots \cite{jumps_in_dots}.  These (stochastic) time records are not captured by the QME, instead we use an unravelled representation in terms of a {pure} density matrix {$\psi_t$} (the conditional state), which has its own stochastic evolution \cite{Belavkin1990,Dalibard1992,Gardiner1992,Carmichael1993AnOS}. 

We use angle brackets $\langle \cdot \rangle$ to indicate averages over the (stochastic) unravelled dynamics.  Note that $\langle \psi_t \rangle = \rho(t)$, which follows the QME \eqref{Lindblad_equation} so the time-dependent density matrix can be obtained as an average of $\psi_t$ over the quantum trajectories.
Since $\psi_t$ is pure, one can always write $\psi_t = \ket{\psi_t}\bra{\psi_t}$ and work with the wavefunction $\ket{\psi_t}$ instead of the density matrix. 
In the following we use $\psi_t$ and $\ket{\psi_t}$ as interchangeable representations of the conditional state, for ease of writing.

In quantum reset processes~\cite{Carollo2019}, 
all jump operators are of rank 1:
\beq
\label{equ:jump-reset}
J_k = \sqrt{\kappa_k} |\phi_k\rangle \langle \xi_k |
\eeq
where $|\phi_k\rangle$ is the reset state (i.e., the jump destination), while $\kappa_k$ and $\langle\xi_k|$ parameterise the jump rate. 
Such models are widespread in recent studies \cite{Thingna2016,Gegg2016,Menczel2020,Snizhko20,Viana2022,Carisch2023}. Repeated global projective measurements at fixed rate also naturally results in reset dynamics, where the reset states $|\phi_k\rangle$ correspond to the eigenstates of the measured observable \cite{Lami2024}.
Quantum trajectories are particularly simple for quantum reset processes~\cite{Carollo2019}.
(Note that quantum reset models are distinct from the `stochastic resetting' considered in \cite{Mukherjee_2018,stochastic_resetting,Kulkarni2023}, in which the trajectory is reset with a fixed rate, independent of the current quantum state.)
It is convenient to define 
\beq
G = -iH_\text{eff}, 
\eeq
where $H_\text{eff} = H -\frac{i}{2}\sum_k J_k^\dagger J_k$ is an effective Hamiltonian which governs the evolution of the conditional state between jumps. If the system jumps by operator $J_j$ at time $t$, the probability that it does not jump again before time $t+\tau$ is the survival probability
\begin{equation}\label{survival_probability}
    S_j(\tau)=\Tr[e^{G\tau} \phi_j e^{G^\dagger\tau}] .
\end{equation}
Moreover, if the system does survive until time $t+\tau$, its conditional state is then $\psi_{t+\tau}=\psi_j(\tau)$
with 
\begin{equation}
\psi_j(\tau) = \frac{e^{\tau G} \phi_j e^{\tau G^\dag}}{S_j(\tau)}  \; .
\label{equ:cond-sm}
\end{equation}
In this state, the rate to jump by operator $J_k$ is 
\begin{align}\label{sm_jump_rate}
w_{jk}(\tau) &= \Tr[J_k\psi_j(\tau) J_k^\dag] 
\nonumber\\ & 
= \kappa_k\langle\xi_k|\psi_j(\tau) |\xi_k\rangle.
\end{align}
The quantum jump Monte Carlo method allows these quantum trajectories to be generated~\cite{Carmichael1993AnOS,Daley_2014}; they can be described mathematically as piecewise deterministic processes, whose stochastic simulation is discussed in Section 7.1 of~\cite{Breuer_and_Petruccione}.

As this rate depends explicitly on the time $\tau$ since the last jump, the sequence of jumps cannot be described by a Markov process.  Instead, it is an example of a semi-Markov process~\cite{Maes2009semi,Carollo2021}.

This observation -- that the conditional state \eqref{equ:cond-sm} only depends on the last jump $J_j$ and the time $\tau$ since this last jump --  allows a simplified analysis of quantum trajectories in quantum reset processes. Instead of following the conditional state itself, we can follow the evolution of $j,\tau$, from which $\psi_t$ can be easily reconstructed.  {Note that the steady-state probability distribution for $\psi_t$ -- which we refer to in the following as the invariant measure -- is entirely supported on states of the form \eqref{equ:cond-sm}.}

\subsection{Classical metastability in trajectories, and the committor}\label{sec:metastable_classical_systems}

In the following, we exploit some established methods from metastability in classical systems, including transition path theory~\cite{E2006,Metzner2009,E2010}.  We briefly summarise these ideas for classical Markov processes on finite configuration spaces, which may be discrete or continuous. 

The central idea is that 
typical trajectories of  metastable systems relax quickly into one of their metastable phases, followed by rare transitions between them: we formalise this notion below.  Anticipating the connection to quantum systems, the associated fast and slow time scales are denoted by $\tauf$ and $\taumet$ respectively.  For simplicity, we focus on the case of two metastable {phases} which we denote as $A$ and $B$.  The extension to more {than two phases} is straightforward.

To identify the phases, we define ``core'' sets of configurations, which are ${\cal S}_A$ for phase $A$, and ${\cal S}_B$ for phase $B$.  (The theory is independent of the specific choice of these sets, as long as certain constraints are met, see below.)  Then any stochastic trajectory can be partitioned into phases, as follows: Let $\chi^A_t=1$ 
if ${\cal S}_A$ was visited more recently than ${\cal S}_B$ and $\chi^A_t=0$ otherwise; 
similarly 
$\chi^B_t=1$ if ${\cal S}_B$ was visited more recently than ${\cal S}_A$ and  and $\chi^B_t=0$ otherwise.  Then if $\chi^A_t=1$ we say that the system is in phase $A$ and similarly if $\chi^B_t=1$ then it is in phase $B$.  If $\chi^A_t=\chi^B_t=0$ then neither core set has been visited during the whole trajectory and we say that the system is not in either phase.

Within this setting, we also define the committor:
For any configuration $x$ outside the core sets, the committor to phase $A$ is
the \emph{probability that a trajectory starting at $x$ hits set ${\cal S}_A$ before set ${\cal S}_B$.} This committor is denoted by $C_A(x)$.  If $x$ is in set ${\cal S}_A$ then define $C_A(x)=1$ and similarly for $x$ in ${\cal S}_B$ then $C_A(x)=0$.   Obviously $C_A(x)+C_B(x)=1$.
The committor $C_A(x)$ can always be estimated numerically by generating many stochastic trajectories starting from $x$, and measuring the fraction that hit ${\cal S}_A$ before ${\cal S}_B$, although this may be computationally expensive.  (In the models considered here, more efficient methods are available, see below.)

The definition of the committor applies for any sets ${\cal S}_{A,B}$.  However, trajectories of metastable systems have two essential properties within this setting: (i) for any initial state, the system almost surely relaxes into one of the phases, on a fast time scale of order $\tauf$; (ii) the residence times within the phases are long, of order $\taumet$.
These features depend weakly on the specific choices of the core sets, as long as they are representative of the configurations explored within the phases.

Now assume as in Sec.~\ref{sec:metastability} that metastability in this classical system is controlled by a parameter $\epsilon$, such that $(\tauf/\taumet)\to 0$ as $\epsilon\to0$. 
We will analyse the committor in this limit
\beq
\commet_A(x) = \lim_{\epsilon\to0} C_A(x) \; .
\label{equ:comstar}
\eeq 
For metastable systems with properties (i,ii) above, it follows that while the committor is defined in terms of hitting times of the core sets ${\cal S}_{A,B}$, 
 {it can also be computed as the probability that $\chi^A_t=1$ for a time $t$ with $\tauf\ll t \ll \taumet$, that is 
 \beq
 \commet_A(x) = \lim_{t\to\infty} \lim_{\epsilon\to0} {\rm Prob}(\chi_t^A=1 | x_0=x) \; .
 \label{equ:comm-prob}
 \eeq 
 
Finally, we identify the \emph{basin of attraction} of phase $A$ as the set of states $x$ with $\commet_A(x)=1$, with a corresponding basin for phase $B$. 
Hence, {for $\epsilon=0$,} trajectories started from within the basins relax quickly into their corresponding phases.\footnote{%
These basins are defined asymptotically as $\epsilon\to0$. For finite  $\epsilon$, it may be useful to define a larger basin as the set of configurations  $x$ with $C_A(x)>1-\delta$ for some small parameter $\delta$, but the asymptotic definition is sufficient for our purposes.}
There are generically configurations $x$ with $\commet_A(x)\neq0,1$, which are in neither basin.  Trajectories started from these configurations may relax into either phase, with finite probability.
 }

\section{Three-State Model}\label{sec:three_state_models}

\subsection{Model, and metastable phenomenology}
\label{sec:three_state_intro}

A canonical model of an open quantum system is the three-state `V' shaped model depicted in Fig. \ref{fig:3_state:sys_diagrams}, which has been well studied in a range of contexts \cite{Itano_1990,thermodynamics_quantum_jump_trajectories,derkacz2006,lesanovsky2013,qd_v} and is known to exhibit metastability \cite{Plenio_Kinght,zoller1987,barkai2004,Macieszczak2016}. This model can be realised experimentally as states in a cavity \cite{catch_reverse_quantum_jump}, energy levels of an atom \cite{Itano_1990,v_exp_atom} and in quantum dots \cite{qd_v}. 

Consider a quantum system with three physical states $\ket{a}$, $\ket{b}$, $\ket{c}$; its Hamiltonian and single jump operator are
\begin{align}
    H &= \Omega_1\left(\ket{a}\!\bra{b}+\ket{b}\!\bra{a}\right)+\Omega_2\left(\ket{a}\!\bra{c}+\ket{c}\!\bra{a}\right)     \nonumber \\
    J_1 &= \sqrt{\kappa_1}\ket{a}\!\bra{b},
    \label{equ:3_state_model}
\end{align}
from which we see that the system is a quantum reset process, {with a semi-Markov representation as depicted in Fig.~\ref{fig:3_state:sys_diagrams}}. Physically, it is natural to think of $\Omega_1,\Omega_2$ as coherent driving terms while $\kappa_1>0$ is a matrix element for spontaneous photon emission.  
Metastability is relevant for small driving frequency $\Omega_2$, specifically
\beq
|\Omega_2| \ll |\Omega_1|,\kappa_1.
\label{equ:meta3}
\eeq
To analyse metastability in the framework of Sec.~\ref{sec:metastability}, we therefore take $\Omega_2=\epsilon\omega_2$ and take $\epsilon\to0$ at fixed $\omega_2,\Omega_1,\kappa_1$.

It is convenient to change basis, defining $\ket{0} = \ket{a}$, $\ket{1} = i\ket{b}$, $\ket{2} = i\ket{c}$.  In this basis, the matrix elements of $G$ are all real and the reset state is $\ket{0}$.  It follows that the steady state of the unravelled dynamics is restricted to conditional states with real matrix elements.

The spectrum of the QME has $\lambda_1=0$ (as it must); in the metastable regime \eqref{equ:meta3}, we explain in Appendix~\ref{app:lind-3state} that it has one small (real) eigenvalue $\lambda_2=O(\epsilon^2)$
that is well-separated from the rest [which are $O(1)$].  Hence, the QME dynamics has fast relaxation to \eqref{equ:rhoAB-meta} on timescale $\tauf = O(1)$, followed by slow relaxation to $\rho_\textsc{ss}$ on timescale $\taumet=O(\epsilon^{-2})$, as described by~\eqref{MS_approx}.  This is shown in Fig. \ref{fig:3_state:QME_dynamics}.

\begin{figure}
    \centering
    \includegraphics[width=0.48\textwidth]{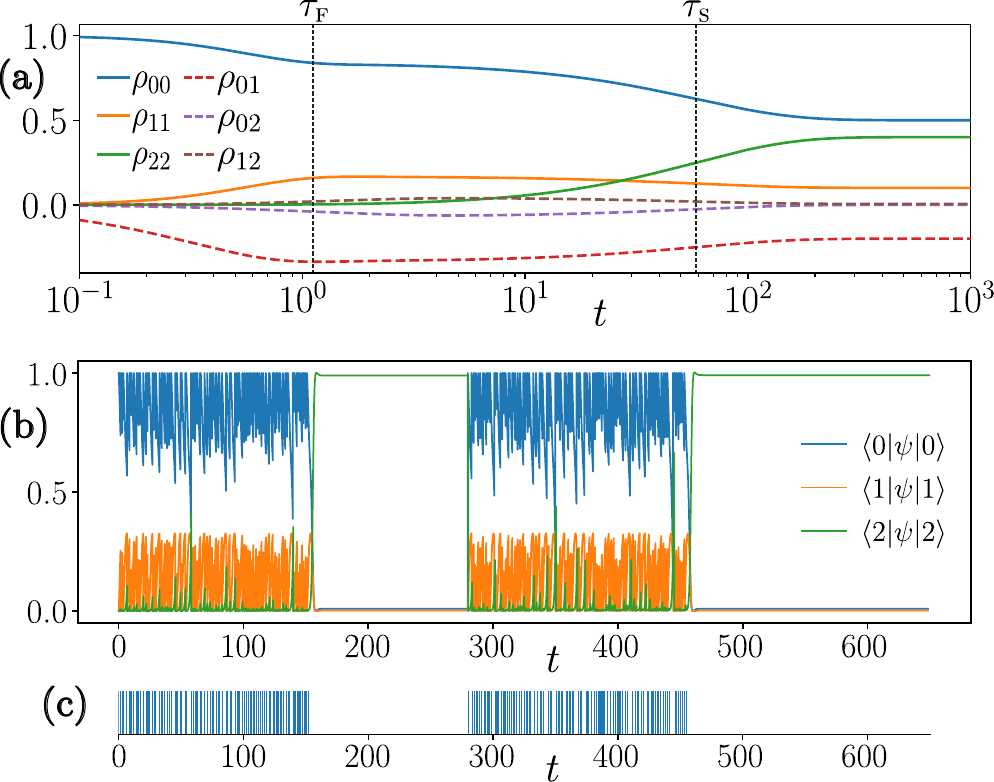}
    \caption{Three-state model. {(a)~QME dynamics from initial state $\ket{0}\!\bra{0}$ with $\Omega_1=1$, $\Omega_2 = 0.05$, $\kappa=4$ and $\rho_{ij}=\bracketmatrix{i}{\rho(t)}{j}$. (b)~Corresponding example unravelled trajectory.
    (c) Measurement record for (b), where each vertical line denotes a jump.}}
    \label{fig:3_state_trajectories}
    \phantomlabel{(a)}{fig:3_state:QME_dynamics}
    \phantomlabel{(b)}{fig:3_state:unravelled_trajectory}
    \phantomlabel{(c)}{fig:3_state:jump_record}
    \phantomlabel{(a,b)}{fig:qme+trajectory_3_state}
    \phantomlabel{(b,c)}{fig:3_state:unravelled_trajectory+measurement_record}
\end{figure}

A representative quantum trajectory and its corresponding measurement record is shown in Figs.~\ref{fig:3_state:unravelled_trajectory+measurement_record}, obtained from a single quantum jump Monte Carlo simulation of the dynamics described in Sec.~\ref{sec:semi-Markov} \cite{Breuer_and_Petruccione,quantum_noise}. It consists of alternating ``bright'' and ``dark'' periods, which correspond to two metastable phases which we label as $B$ (bright) and $D$ (dark).  These correspond to the phases $A,B$ anticipated in Secs.~\ref{sec:metastability} and \ref{sec:metastable_classical_systems}. The dark phase consists of long time periods where the conditional state is ``shelved'' with $\ket{\psi_t} \approx \ket{2}$, and no jumps take place.  Within the bright phase the state $\psi_t$ fluctuates; the jump rate is of order unity so jumps occur frequently; and $\langle{2}|{\psi_t}|{2}\rangle$ remains small.  The large rate of quantum jumps corresponds to frequent photon emissions, hence the name ``bright''.

\subsection{Connections between trajectory and QME dynamics}
\label{sec:connection_trajectories_QME}

Figs.~\ref{fig:qme+trajectory_3_state} highlight the different information that is available in the averaged (QME) dynamics for $\rho(t)$ and the quantum trajectories for the conditional state $\psi_t$. One always has $\rho(t)=\langle\psi_t\rangle$ on average, but \emph{typical} states $\psi_t$ are not close to $\rho(t)$.  This situation is generic for metastable systems with intermittent trajectories.

To understand the relationships between the averaged dynamics and the quantum trajectories, we consider a perturbative analysis about $\epsilon=0$~\cite{Macieszczak2016,classical_metastability_in_quantum}.
In the degenerate case $\epsilon=0$, the Hilbert space is broken into two subspaces: one contains the levels $\ket{0}$, $\ket{1}$ and corresponds to the bright phase; the other is just level $\ket{2}$. As in \eqref{equ:rhoAB-meta}, an initial density matrix with support on both components relaxes quickly under the QME dynamics, and arrives in a linear combination of the two EMS:
\beq
\rho(t) \approx p_B \rho_B + p_D \rho_D 
\label{equ:MS-BD}
\eeq
for $\tauf \ll t \ll \taumet$,
where $\rho_B,\rho_D$ represent the bright and dark phases respectively, and are supported on the two subspaces.

Turning to the unraveled dynamics with $\epsilon=0$, every trajectory relaxes quickly into either the bright phase or the dark phase: in the notation of Sec.~\ref{sec:metastable_classical_systems}, this means that either $\chi_t^B=1$ or $\chi_t^D=1$ after a short time of order $\tauf$, after which these variables do not change. 
(In Sec.~\ref{sec:metastable_unravelled} below, we establish some general relationships between quantum trajectories and the QME, showing that 
the probabilities of relaxing into each phase are $p_B,p_D$.)
After this fast relaxation, the key point is that $\psi_t$ is typically close to either $\rho_B$ or $\rho_D$, depending on the phase into which the system relaxed. This differs from the average state in \eqref{equ:MS-BD} as long as both $p_B$ and $p_D$ are non-zero.

On increasing $\epsilon$ from zero, the system becomes ergodic: the QME dynamics now has two-step relaxation to a unique steady state and the trajectories show rare transitions between the metastable phases (recall Fig.~\ref{fig:3_state_trajectories} and the associated discussion).  On these long time scales $t\sim \taumet$, the coefficients $p_B,p_D$ in~\eqref{equ:MS-BD} acquire time-dependence according to \eqref{MS_approx}: we have $m=2$ so this is well-described by a single exponential with (real) rate $\lambda_2=O(\epsilon^2)$.  For consistency with the unravelled trajectories, this $\lambda_2$ must be the rate for the rare transitions between the phases: this is an additional connection between quantum trajectories and QME dynamics \cite{meta_ising,classical_metastability_in_quantum}.  Finally, note that in addition to this (non-perturbative) restoration of ergodicity on taking $\epsilon>0$, the EMS $\rho_B,\rho_D$ are affected (perturbatively) by $\epsilon$, acquiring weak coherences, for example $\bracketmatrix{1}{\rho_\textsc{ss}}{2}=O(\epsilon)$.

\subsection{Jumpless trajectory}
\label{sec:jumpless-3state}

To analyse the system further, we exploit the fact that it is a quantum reset process. This means that when a jump occurs, the subsequent evolution is determined by the effective Hamiltonian according to \eqref{equ:cond-sm}, with $\phi_j = \ket{0}\bra{0}$.  The system follows this evolution until the next jump occurs, after which the process repeats.
Similarly to \eqref{equ:cond-sm}, this means that
\beq
\ket{\psi_{\tau}} \propto {\rm e}^{\tau G} |0\rangle
\label{equ:cond3}
\eeq
where $\tau$ is the elapsed time since the last jump, and the constant of proportionality is fixed by normalisation.  
This equation describes the time evolution of the quantum state after a jump, under the assumption that no further jumps takes place.  Hence we refer to it as the ``jumpless trajectory''.

Of course, this (deterministic) jumpless trajectory is not typical for the (stochastic) unravelled dynamics.  Instead, typical trajectories of the unravelled dynamics can be constructed by piecing together segments of the jumpless trajectory, each of which starts from the reset state, interspersed by jumps.  Since there is a single reset state, the invariant measure for $\psi_t$ is entirely supported on this jumpless trajectory.


\begin{figure}
    \centering
    \includegraphics[width=85mm]{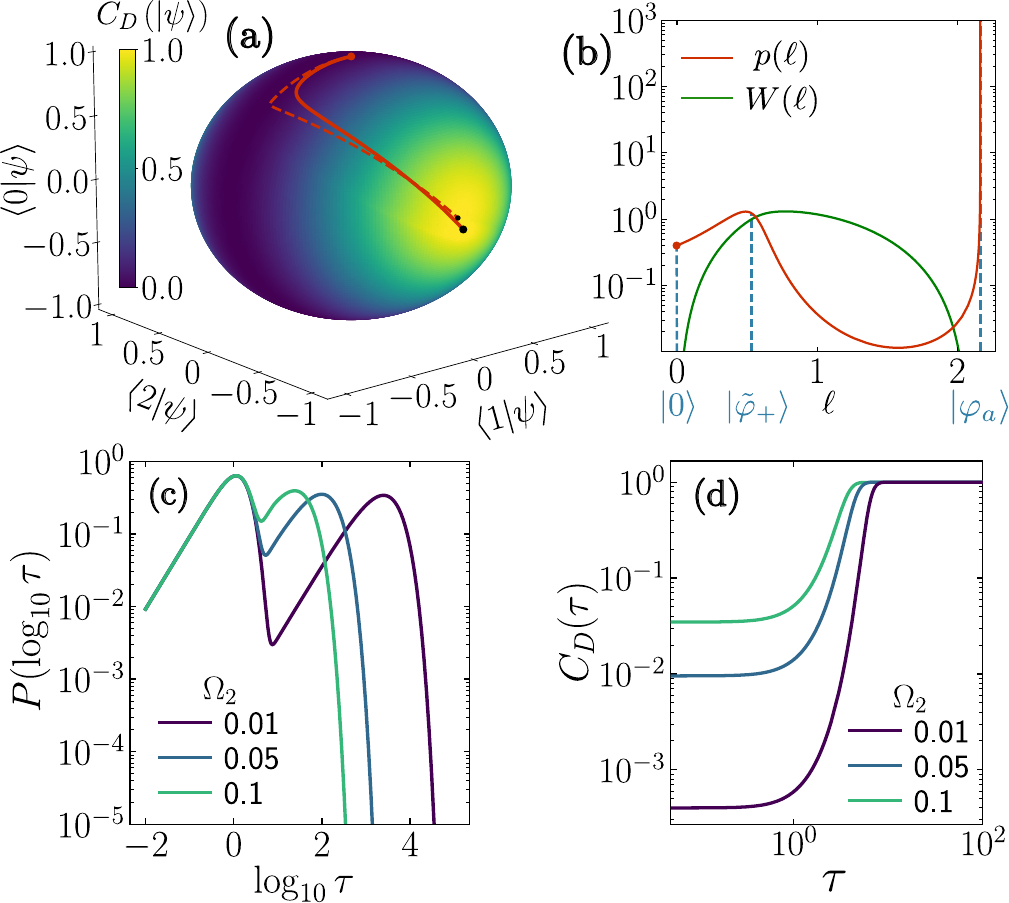}
    \caption{Three-state model. (a) Trajectories of the (real) wavefunction $\ket{\psi_t}$ can be plotted on a sphere.  Red lines are the jumpless trajectory for $\Omega_2=0.05$ (solid) and $\Omega_2=0.0001$ (dashed) with 
    $\kappa=4$, $\Omega_1=1$.  The shading shows the committor to the dark phase.
    (b) Steady-state probability density $p(\ell)$ for the distance $\ell$ travelled along the jumpless trajectory. Here $\Omega_2=0.05$.  We also show the corresponding jump rate $w_{11}(\ell)$.
    (c) Probability density for the time since the last jump, within the steady state.
    (d) Committor to the dark phase as a function of the time since the last jump.
    All panels take $\Omega_1=1$, $\kappa_1=4$ as in Fig. \ref{fig:3_state_trajectories}.}
    \label{fig:3_state_panel2}
    \phantomlabel{(a)}{fig:3_state:sphere_plot}
    \phantomlabel{(b)}{fig:3_state:length_IM}
    \phantomlabel{(c)}{fig:3_state:SM_IM}
    \phantomlabel{(d)}{fig:3_state:SM_committor}
    \phantomlabel{(a,d)}{fig:a_d_3_state}
\end{figure}

\newcommand{\rrr}{w}

In our chosen basis, $G$ is a (non-symmetric) $3\times3$ matrix with real elements so the elements of $\ket{\psi_{\tau}}$ are also real.  
Hence, $\ket{\psi_{\tau}}$  can be represented as a point on the surface of a sphere.
{For simplicity, we focus our discussion on the parameter regime $\kappa_1\geq 4\Omega_1$, the alternative case is discussed in Appendix~\ref{appendix:3_state}.}
Fig.~\ref{fig:3_state:sphere_plot} shows the resulting jumpless trajectory for illustrative parameters, representative of the metastable regime~\eqref{equ:meta3}.  The north pole of the sphere corresponds to the reset point: the trajectory extends away from this point and eventually converges towards to a stationary point.  From \eqref{equ:cond3}, this is the eigenvector of $G$ with largest real part, which we denote by $\ket{\varphi_a}$.
In addition, we have from~\eqref{sm_jump_rate}, the jump rate for states on the jumpless trajectory is 
\beq
\rrr_{11}(\tau) = \kappa_1\langle 1|\psi_1(\tau)|1\rangle \; .
\label{equ:rrr}
\eeq

Using (\ref{equ:cond3},\ref{equ:rrr}) together is helpful for understanding the unravelled dynamics of this model.  
Some 
properties of the jumpless trajectory are computed in Appendix~\ref{app:G-3state}, we summarise here the main results and their implications:
 We have $\ket{\varphi_a}=\ket{2}+O(\epsilon)$. 
The eigenvector of $G$ with the second largest real part is denoted by $\ket{\varphi_+}$.  Starting from the reset point $\ket{0}$, the jumpless trajectory
evolves quickly (on a time scale of order unity) to
a state {$\ket{\tilde{\varphi}_+}=\ket{\varphi_+}+O(\epsilon)$}:
this corresponds to the point where the trajectory abruptly changes direction in Fig.~\ref{fig:3_state:sphere_plot}, which we refer to as the ``elbow''.
The jump rate $\rrr_{11}(\psi)$ is large (of order $1/\tauf$) between the reset point and the elbow, so the typical dynamical behaviour involves repeated rapid motion from $\ket{0}$ towards $\ket{\Tilde{\varphi}_+}$, interspersed with frequent jumps back to $\ket{0}$. This is the characteristic behaviour of the bright phase, the corresponding average activity is $\kappa_1\langle1|\rho_B|1\rangle=O(1)$. 

{To understand the transition to the dark phase, note that the jumpless trajectory slows down near the elbow: Appendix~\ref{app:G-3state} shows that it remains close to $\ket{\Tilde{\varphi}_+}$ for a time of order $\tauf\log(1/\epsilon)$. On longer times, it evolves towards $\ket{\varphi_a}$ where the jump rate is $\kappa_1\langle 1|\varphi_a|1\rangle=O(\epsilon^2)$.  This corresponds to the dark phase: we see that the transition mechanism from bright to dark is a continuous evolution along the jumpless trajectory.  Such events are rare because the jump rate is large near the elbow and the state remains there a long time:  Appendix~\ref{app:G-3state} shows that the probability to pass the elbow before jumping is $O(\epsilon^2)$.  By contrast, the transition from dark to bright occurs by a jump back to $\ket{0}$: this is rare because of the small jump rate.}
The two dynamical regimes can both be seen in the trajectory Fig.~\ref{fig:3_state:unravelled_trajectory}: it starts in the bright phase and visits the dark phase twice.

The above analysis illustrates that the steady state of the unravelled dynamics only visits states on the jumpless trajectory, as expected for a quantum reset process with a single jump operator. Fig.~\ref{fig:3_state:length_IM} shows how the steady state probability density can be parameterised in terms of the distance $\ell$ travelled along the jumpless trajectory.
Here, the distance between two nearby points $\psi$ and $\psi+d\psi$, denoted by $d\ell$, is given by the trace distance of the corresponding pure states $d\ell = \frac{1}{2}\|d\psi\|_1 = \frac{1}{2}\sum_i|\lambda_i|$, where $\lambda_i$ are the eigenvalues of $d\psi$ which are real \cite{Wilde2016}. Hence, the distance along the jumpless trajectory at time $\tau$ since the last reset is given by $\ell(\tau)=\frac{1}{2}\int_0^\tau \|\dot\psi(t)\|_1 dt$, where $\dot\psi(\tau)$ is the time derivative of the jumpless trajectory \eqref{equ:cond-sm}. The steady state probability density along the jumpless trajectory is given by $p(\ell)=p(\tau)d\tau/d\ell$, where $p(\tau)$ is the steady state probability density parameterised by time since the last reset.

The probability shows two peaks, corresponding to the two phases.  The bright phase forms a broad peak centred near $\ket{\Tilde{\varphi}_+}$ while the dark phase is a sharp peak near $\ket{\varphi_a}$.
For large $\tau$, both $p(\tau)$ and $d\ell/ d\tau$ decay exponentially as $\ln(p(\tau))=O(\epsilon)\tau$ and $\ln(d\ell/ d\tau)=O(1)\tau$ respectively. The faster decay of $d\ell/d\tau$ results in $p(\ell)$ diverging as $\ell\rightarrow\ell_\text{max}$, i.e. as $|\psi\rangle$ approaches $|\phi_a\rangle$.

\subsection{Committor and semi-Markov analysis}

The shading in Fig.~\ref{fig:3_state:sphere_plot} shows the committor $C_D(\psi)$.  Recalling  Sec.~\ref{sec:metastable_classical_systems}, this is the probability that a trajectory started in state $\psi$ reaches the dark phase before the bright phase. 
For any stochastic dynamics, the committor can always be estimated numerically by running many stochastic trajectories starting from $\psi$.  However,  Sec.~\ref{sec:committor_for_reset} derives a generic formula for the committor in quantum reset processes, which allows more efficient numerical estimation. The results of Fig.~\ref{fig:3_state:sphere_plot} use this more efficient method.

It is convenient to take the core set of the bright phase ${\cal S}_B$ as the single point $|0\rangle$, so the hitting time for this set is the time of the first jump.  For the dark phase we take ${\cal S}_D$ as a small ball around $|\varphi_a\rangle$.
As discussed in Sec.~\ref{sec:metastable_classical_systems}, the committor can be used to identify the two phases within the Hilbert space: $C_D$ is large when $\psi$ is close to the dark phase, and small when it is far away.

As explained in Sec.~\ref{sec:semi-Markov}, the jumps in this quantum reset process are semi-Markov: as long as at least one jump has taken place, the state $\ket{\psi_t}$ is restricted to the jumpless trajectory in Fig.~\ref{fig:3_state:sphere_plot}, and its position on this line only depends on the time $\tau$ since the last jump.  {This means that after the first jump has taken place,} the unravelled dynamics can be reduced to a reset process for the random variable $\tau\in[0,\infty)$, or equivalently for $\log\tau$.  To illustrate this, Fig.~\ref{fig:3_state:SM_IM} shows the steady state probability distribution of $\log\tau$.   It shows two peaks, which correspond to the two metastable phases.

Within the semi-Markov representation, the committor becomes a function of $\tau$, 
shown in
Fig.~\ref{fig:3_state:SM_committor}.  Recalling that the committor $C_D$ is large for systems in the dark phase and small for those in the bright phase, we see that 
the bright phase corresponds to small $\tau$ and the dark phase to large $\tau$.
The semi-Markov jump rate $w_{11}(\tau)$ is shown in Fig.~\ref{fig:3_state:length_IM}, parameterised as a function of distance along the jumpless trajectory.  Metastability in this semi-Markov representation arises because $w_{11}(\tau)$ is large when $\tau$ is small, but decreases strongly for large $\tau$: this encapsulates the effect of shelving in the dark state.
These results in the semi-Markov representation illustrate the advantage of the quantum reset process for analysis of quantum trajectories: a generic three-state quantum system has a dynamical evolution of $\ket{\psi_t}$ in a 3-dimensional Hilbert space, but the invariant measure and the basins of attraction of the phases can be analysed via a one-dimensional stochastic process for $\tau$.

\subsection{Two-jump variant of three-state model}
\label{sec:3_state_2_jump}

The three-state model considered thus far illustrates several aspects of metastability but  
it also has special features, due to its simplicity.  In particular, {as $\epsilon\to0$ the dark metastable phase has no activity at all.}

To illustrate a more generic situation we modify the model by adding an additional jump operator [see Fig.~\ref{fig:two_jump:sys_diagrams}]:
\beq
    J_2 = \sqrt{\kappa_2} |2\rangle\langle 2|,
\eeq
{Recalling \eqref{sm_jump_rate},
this operator manifests in the unravelled dynamics as a jump into state $\ket{2}$ with rate {$\rrr_{j2}(\tau) = \kappa_2\langle 2|\psi_j(\tau)|2\rangle $} that depends on the destination $j$ of the last jump, as well as the time $\tau$ since that jump.
 At the level of the QME, the operator $J_2$ corresponds to dephasing within the dark phase. We take $\Omega_2 = \epsilon\omega_2$ as before, and $\kappa_2=O(1)$. This choice ensures that the model is still metastable as $\epsilon\to0$.}

{Figs.~\ref{fig:3_state_2_jump:trajectory+record} depict a representative trajectory and corresponding measurement record.} The behaviour is similar to Fig.~\ref{fig:3_state_trajectories}, except that the ``dark phase''  now includes fluctuations of the quantum state, with a finite rate of internal jumps whose statistics are approximately Poissonian {with rate $\kappa_2\langle 2|\varphi_a|2\rangle$}.  (Despite this fact, we continue to refer to it as the ``dark phase'', to aid comparison with previous Sections.)

The system is still a quantum reset process so its quantum trajectories have a semi-Markov representation, {as depicted in Fig.~\ref{fig:two_jump:sys_diagrams}}.  However, it is slightly more complex than the original 3-state model because
there are two different jumpless trajectories, one starting from each reset point (which are $|0\rangle$ and $|2\rangle$). The resulting situation is illustrated in Fig.~\ref{fig:3_state_2_jump:sphere_plot}.
Both jumpless trajectories end at the same point: this is the dominant eigenvector of $G$, which we continue to denote by $|\varphi_a\rangle$.  
As $\epsilon\to0$, we have $|\varphi_a\rangle \approx |2\rangle$, as in the original three-state model.
We emphasize that the invariant measure for $\psi_t$ is fully supported on these two jumpless trajectories: this is the essential simplification that is available for quantum reset processes.

\begin{figure}[t]
    \centering
    \includegraphics[width=85mm]{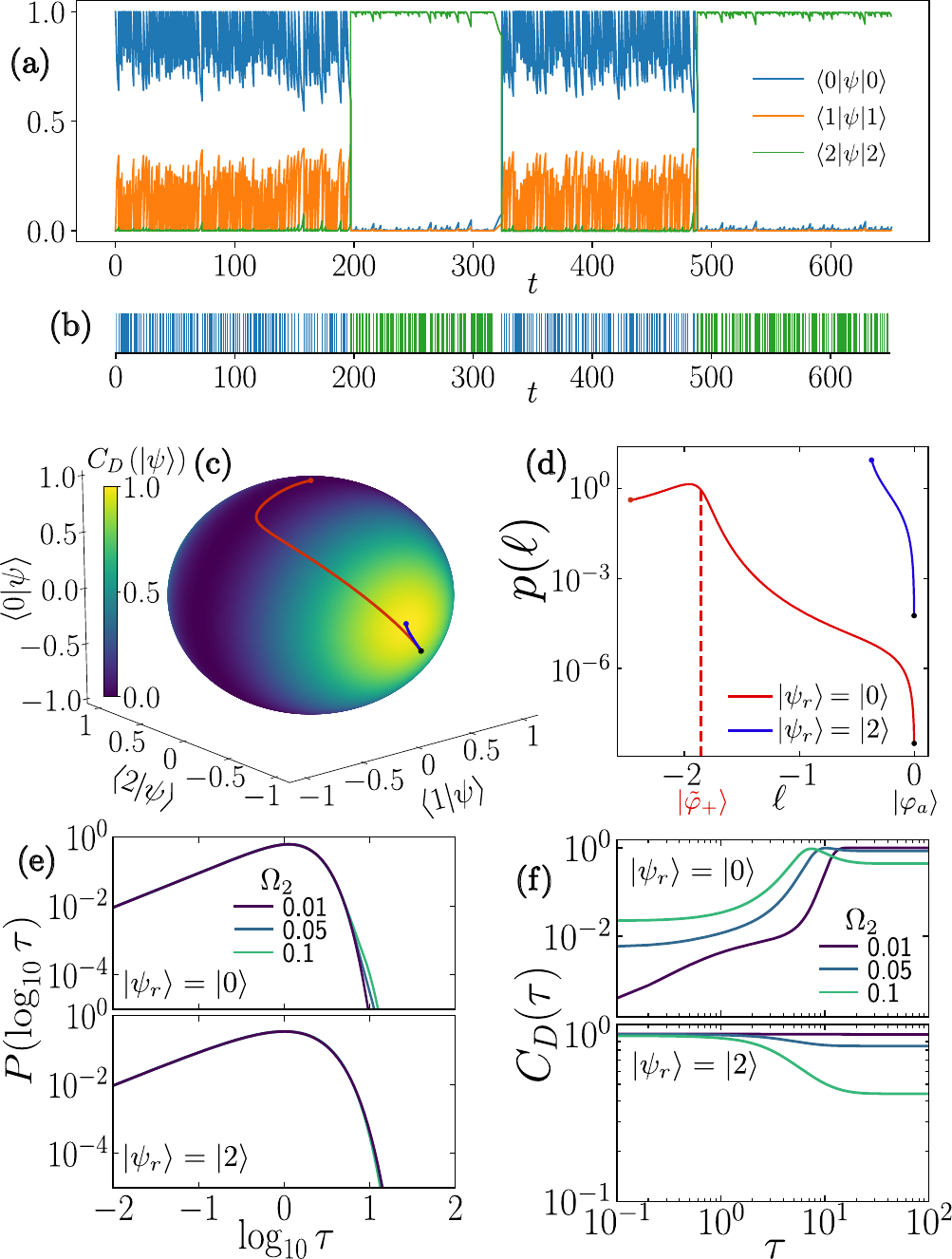}
    \caption{
    {Two-jump variant of three-state model. (a)~Example unravelled trajectory, from initial state $|0\rangle\langle 0|$. (b)~Measurement record for (a) with jumps $J_1$ (blue) and $J_2$ (green). 
    (c) Jumpless trajectories from the reset states, $|0\rangle$ (red) and $|2\rangle$ (blue) to the asymptotic state (black dot). The shading shows the committor to the  dark phase. (d) Probability density for the steady state, as a function of distance along the jumpless trajectories, from each reset state denoted by $|\psi_r\rangle$. (e)~Probability density for the time since the last jump, within the steady state, from each reset state. (f) Committor to the dark phase as a function of the time since the last jump, from each reset state, calculated using \eqref{committor_from_jumps}. All panels take $\Omega_1=1$, $\kappa_1=4$, $\kappa_2=1$, with $\Omega_2=0.05$ in (a-d).}}
    \label{fig:3_state_2_jump_panel}
    \phantomlabel{(a)}{fig:3_state_2_jump:unravelled_trajectory}
    \phantomlabel{(b)}{fig:3_state_2_jump:measurement_record}
    \phantomlabel{(c)}{fig:3_state_2_jump:sphere_plot}
    \phantomlabel{(d)}{fig:3_state_2_jump:length_IM}
    \phantomlabel{(e)}{fig:3_state_2_jump:SM_IM}
    \phantomlabel{(f)}{fig:3_state_2_jump:SM_committor}
    \phantomlabel{(a,b)}{fig:3_state_2_jump:trajectory+record}
    \phantomlabel{(c,f)}{fig:c_f_3_state_2_jump}
\end{figure}

To characterise the stochastic trajectory dynamics, we again consider the probability distribution for the distance travelled along the jumpless trajectories [Fig.~\ref{fig:3_state_2_jump:length_IM}]. Since each jumpless trajectory ends at the same asymptotic point, we now measure the length backwards from that point, so we take $\ell$ to be negative, with the final point ($\ell=0$) corresponding to $\ket{\psi_t}=\ket{\varphi_a}$.  
The invariant measure now has contributions from both jumpless trajectories: the probability density for the system to be on trajectory $j$ at a distance $\ell$ from the corresponding reset point is $p_j(\ell)$, with normalisation $\sum_{j=1}^2 \int_{-\infty}^0 p_j(\ell) d\ell = 1$.
One metastable phase is concentrated on each jumpless trajectory.  Hence, in contrast to the bimodal distribution in Fig.~\ref{fig:3_state:length_IM}, each $p_j$ now has a single peak, which corresponds to one of the coexisting phases.

Fig.~\ref{fig:3_state_2_jump:SM_IM} shows the corresponding distributions of the time since the last jump.  {To understand the mechanisms of transition between the phases, observe that the system leaves the dark phase by a jump to $\ket{0}$, as in the original model of~\eqref{equ:3_state_model}.  It may transition to the dark phase either by passing the elbow along the jumpless trajectory (as in the original model) or by a direct jump to $\ket{2}$: the relative probabilities of these two mechanisms depends on $\kappa_2$.}

The existence of two reset states means that the steady state in the semi-Markov representation is now represented as a joint probability distribution for the type of the last jump and the time since this jump.  This means that the invariant measure in Fig.~\ref{fig:3_state_2_jump:SM_IM} consists of two histograms, instead of the single histogram in Fig.~\ref{fig:3_state:SM_IM}.
Nevertheless, the common features of the {two variants of the model} are that they both exhibit intermittent dynamics of trajectories, with fast relaxation into metastable states, and rare transitions between them.

Fig.~\ref{fig:3_state_2_jump:SM_committor} shows the committor in the semi-Markov representation, analogous to Fig.~\ref{fig:3_state:SM_committor}.  The interpretation is that the jumpless trajectory starting from $\ket{2}$ is predominately in the dark phase ($C_D$ is large); the jumpless trajectory starting from $\ket{0}$ starts in the bright phase ($C_D$ is small for small $\tau$) but it crosses over to the dark phase when $\tau$ is large [similar to Fig.~\ref{fig:3_state:SM_committor}].  There are four jump rates $w_{jk}(\tau)$ in the semi-Markov representation (indexed by $j,k$): the important feature is that {the jump rate} into $\ket{2}$ is large in the dark phase and small in the bright phase, and vice versa for {the jump rate} into $\ket{0}$.  This means that after a jump into either reset state, it is overwhelmingly likely that the next jump will return to the same reset state -- this is generic for metastability in reset models where both states have finite activity.

\section{The committor}
\label{sec:metastable_unravelled}

We have seen that the committor is a useful quantity for the analysis of quantum trajectories, especially as a way to identify the distinct metastable phases in Figs.~\ref{fig:3_state:sphere_plot} and~\ref{fig:3_state_2_jump:sphere_plot}. We now derive general properties of this object, which directly aid its computation. In doing so we develop generic connections between the committor (which depends on trajectories) and the quantum master equation (which describes the average dynamics).  These generic results are not restricted to quantum reset models.  
We also derive some additional properties of the committor that hold for the specific case of quantum reset models.

\subsection{Relation of the committor to the QME}\label{sec:relation_committor_QME}

The unravelled dynamics of the
models discussed so far has all the features of classical metastability: a system started in any state $\ket\psi$ typically relaxes quickly into either the bright or the dark phase; it explores that phase quickly and it resides there for a long time, before eventually transitioning into the other phase.  
The fast relaxation is stochastic, the probability to relax into the dark phase is the committor $C_D$.

Now consider
{a general open quantum system} with two-phase metastability. 
After initialisation in pure state $\psi_0$ at time $t=0$, we consider the probability distribution $\mu(\psi,t)$ of the conditional state $\psi_t$.  
Assuming that $\psi_t$ relaxes quickly into one of the phases $A$ or $B$ (such that either $\chi_t^A=1$ or $\chi_t^B=1$ after a time of order $\tauf$),
we have by \eqref{equ:comm-prob} that for $\tauf \ll t \ll \taumet$,
\beq
    \mu(\psi,t) \approx  C_A^*(\psi_0) \mu_A(\psi) + [1-C_A^*(\psi_0)] \mu_B(\psi)
    \label{equ:pAB}
\eeq
where all dependence on the initial condition appears through (the asymptotic value of) the committor $C_A^*(\psi_0)$, and $\mu_{A,B}(\psi)$ are probability distributions for the {conditional state}, within the metastable phases.
The approximate equality appears because we rely on the separated time scales $\tauf \ll t \ll \taumet$.

Recalling that $\rho(t)=\langle \psi_t\rangle$, averaging $\psi$ against the distribution in \eqref{equ:pAB} yields
\beq
    \rho(t) \approx  C_A^*(\psi_0) \rho_A +[1-C_A^*(\psi_0)] \rho_B
    \label{equ:rhoAB-comm}
\eeq
where $\rho_{A},\rho_{B}$ are the averages of $\psi$ with respect to $\mu_A,\mu_B$. Since \eqref{equ:rhoAB-meta} holds in the same asymptotic limit, it is natural to compare these two results, which suggests that we identify $p_A$ with $C_A(\psi_0)$, and similarly for phase $B$.  Noting that (\ref{equ:rhoAB-meta},\ref{equ:rhoAB-comm}) are approximate equalities that become accurate in the limit $\epsilon\to0$, one finds
\beq\label{equ:CAP}
    \commet_A(\psi) = \Tr [ {P}_A^* \psi ], \quad \commet_B(\psi) = \Tr [ {P}_B^* \psi ],
\eeq
with ${P}_{A,B}^* = \lim_{\epsilon\to0} {P}_{A,B}$.

Eq.~\eqref{equ:CAP} -- which is generic for unravelled systems with two-phase metastability -- is important as a connection between the QME and trajectory representations of metastability.  
For example if $P_A^*\ket{\psi}=\ket{\psi}$ then $C_A^*(\psi)=1$ so this state is in the basin of attraction of phase $A$, and quantum trajectories started in $\ket{\psi}$ will (typically) relax quickly into that phase.  A similar property holds if $P_B^*|\psi\rangle=|\psi\rangle$, in which case $\ket{\psi}$ is in the basin of attraction of phase $B$ (and $P_A^*|\psi\rangle=0$).  For initial (pure) states $\psi$ which overlap with both phases, the committor is intermediate: they relax to one phase or the other with probability $\commet_{A,B}(\psi)$. (The extension of these results to more than two phases is straightforward.)

An interesting feature of \eqref{equ:CAP} is that the committor is independent of the unravelling (for example, the same formula applies also for homodyne unravellings~\cite{quantum_noise}).  Hence, the basins of attraction of the phases $A$ and $B$ are also independent of the unravelling.  We also emphasise that \eqref{equ:pAB} -- which is the starting point for this analysis -- relies on the \emph{assumption} that the unravelled dynamics relaxes quickly into either phase $A$ or phase $B$.  This certainly holds in the 3-state examples considered here and we expect it to hold in a broad range of examples, but a detailed investigation of the conditions required remains as an interesting  direction for future work.

Having identified these general principles, we summarise their implications for the 3-state models.  Any trajectory which starts in a quantum superposition of the two phases, such as $(\ket{0} + i \ket{2})/\sqrt{2}$, will evolve {on the fast timescale $\tauf$} into either the dark or the bright phase, according to these committor probabilities.
In the limit $\epsilon\to0$, 
the relevant operators are projectors:
$P_D^*=\ket{2}\bra{2}$ and $P_B^*=(\ket{0}\bra{0}+\ket{1}\bra{1})$. These are projections onto the bright/dark subspaces of the model. The committors to each phase -- which are computed directly in Sec.~\ref{sec:committor_for_reset} -- are related to the underlying QME by \eqref{equ:CAP}. Hence as $\epsilon\to0$ they converge to
\begin{equation}
    C_B^*(\psi) = \bracketmatrix{0}{\psi}{0} + \bracketmatrix{1}{\psi}{1}, \quad C_D^*(\psi) = \bracketmatrix{2}{\psi}{2}.
    \label{equ:CB-CD-formulae}
\end{equation}

\subsection{Committor for reset processes}
\label{sec:committor_for_reset}

For reset processes, the core sets ${\cal S}_A,{\cal S}_B$ that appear in the definition of the committor can often be identified with reset states, as they were in Sec.~\ref{sec:3_state_2_jump}. Here we derive a formula for the committor that applies in this situation.  (We note however that this choice of core sets is not possible in all cases: an example is given by the model of Sec.~\ref{sec:three_state_intro}, which has only one reset state.)

For a quantum trajectory starting in  state $\psi_0$ at time $t=0$, the probability that the first jump has destination $\ket{\phi_j}$ and occurs
 in the time interval $[t,t+dt)$ is 
 $p(j,t|\psi_0)dt$, with
\beq
p(j,t|\psi_0) =S(t,\psi_0)w_j(t,\psi_0)
\eeq
where $S(t,\psi_0)=\Tr[e^{Gt}\psi_0 e^{G^\dagger t}]$ is the survival probability up to time $t$ [analogous to \eqref{survival_probability}] 
and 
\beq
w_j(t,\psi_0) = \frac{\Tr[J_j e^{Gt}\psi_0 e^{G^\dagger t} J^\dag_j]}{
\Tr[e^{Gt}\psi_0 e^{G^\dagger t}]}
\eeq
is the jump rate by jump $J_j$ at time $t$ [analogous to 
\eqref{sm_jump_rate}].  
For quantum reset processes we can then use \eqref{equ:jump-reset} to obtain
\beq
p(j,t|\psi_0) = \kappa_j 
\big\langle \xi_j \big\vert e^{Gt}\psi_0 e^{G^\dagger t} \big\vert \xi_j\big\rangle.
\eeq
Hence, the total probability that the first jump after $t=0$ is of type $j$ is 
\beq\label{eqn:marginal}
P(j|\psi_0)=\int_0^\infty dt\, p(j,t|\psi_0).
\eeq
This is an example of a \emph{splitting probability}~\cite{vankampen2007spp}.

Now consider a system with two or more phases, where the core set for each phase $A,B,\dots$ consists of a single reset state,
with corresponding jump operators of the form \eqref{equ:jump-reset}.  Then the committor from state $\psi_0$ to phase $A$ is exactly the probability that a quantum trajectory starting at $\psi_0$ makes its first jump to $\ket{\phi_A}$, which is
\begin{align}\label{eqn:committor_reset_core_sets}
    C_A(\psi_0) =
    \int_0^\infty d\tau\, \kappa_A \big\langle \xi_A \big\vert e^{Gt}\psi_0 e^{G^\dagger t} \big\vert \xi_A\big\rangle .
\end{align}

For example, in the model of Sec.~\ref{sec:3_state_2_jump}, one identifies the jump operator corresponding to jumps into the ``dark'' phase as $J_2 = \sqrt{\kappa_2}\ket{2}\bra{2}$ and \eqref{eqn:committor_reset_core_sets} reduces to
\begin{equation}\label{committor_from_jumps}
    C_D(\psi_0) = \int_0^\infty d\tau\, \kappa_2 \langle 2|e^{G\tau}\psi_0 e^{G^\dagger\tau}|2\rangle,
\end{equation}
with a similar formula for $C_B$ on replacing $\ket{2}\to\ket{1}$ and $\kappa_2\to\kappa_1$.
The committor for this model, calculated from \eqref{committor_from_jumps}, is shown in Figs.~\ref{fig:c_f_3_state_2_jump}.
Note that as $\epsilon\rightarrow 0$, \eqref{committor_from_jumps} converges to \eqref{equ:CB-CD-formulae}.

The result \eqref{eqn:committor_reset_core_sets} has practical implications. It is much more efficient than estimating committors directly by running many random trajectories starting from each state.  Of course one can also use \eqref{equ:CAP} to obtain the committor in the limit $\epsilon\to0$, but \eqref{eqn:committor_reset_core_sets} is useful since it is valid also at finite $\epsilon$, and it also avoids the requirement to diagonalise the quantum Liouvillian operator.  In addition, while \eqref{eqn:committor_reset_core_sets} is not directly applicable to the first model of Sec.~\ref{sec:three_state_models}, a modified version of the formula still applies, where the upper limit on the integral is the time at which the jumpless reaches the core set ${\cal S}_D$ for the dark state.  This last method was used to compute the committor in Figs.~\ref{fig:a_d_3_state}.

\section{Two qubit model}\label{sec:two_qubit_model}

As well as the two-phase metastability considered so far, the
general theory of quantum metastability includes a range of other phenomena including 
noiseless subsystems (NSS) and decoherence free subspaces (DFS)\cite{dfs_review_2014}, which are protected from dissipation and decoherence, and have been proposed as possible candidates for the implementation of quantum information processing \cite{Palma_1996,Lidar1998}.


\subsection{Quantum reset model with metastable DFS}\label{sec:DFS_two_qubit}

As an example of a metastable DFS in a quantum reset process, we 
consider two coupled qubits, with Hamiltonian and jump operators given by
\begin{align}
    H &= \Omega_1\sigma_1^y+\Omega_2\sigma_2^y     \nonumber \\
    J_1 &= \sqrt{\gamma_1}n_1\sigma_2^- 
    \label{DFS_H_J} \\
    J_2 & = \sqrt{\gamma_2}(1-n_1)\sigma_2^+,
    \nonumber
\end{align}
where $\sigma_i^\pm=\frac{1}{2}\left(\sigma_i^x\pm\sigma_i^y\right)$, the subscript denotes the qubit on which the operator acts, $n_i=\frac12(1+\sigma^z_i)$, and we use the single qubit basis $\{\ket{\xup},\ket{\xdown}\}$.  This system is {depicted in Fig.~\ref{fig:dfs_reset:sys_diagrams} and} based on a similar model, proposed in \cite{Macieszczak2016}. It is defined in a basis for which the matrix elements of $G$ and ${\cal L}$ are all real.
The model is a quantum reset process with two reset points, $\ket{\xup\xdown}$ and $\ket{\xdown\xup}$. The semi-Markov structure of the model is illustrated in Fig.~\ref{fig:dfs_reset:sys_diagrams}. Note that this has the same structure as for the two-jump three-state model shown in Fig.~\ref{fig:two_jump:sys_diagrams}, but the differing waiting time distributions of the semi-Markov processes lead to very different qualitative behaviour.

The metastable regime is
\beq
|\Omega_1|, |\Omega_2| \ll \gamma_1, \gamma_2.
\eeq
For consistency with the general formalism of Sec.~\ref{sec:metastability}, we take $\Omega_{1,2}=\epsilon \omega_{1,2}$ with $\gamma_{1,2}>0$ and $\omega_{1,2}$ held constant as $\epsilon\to0$.  In this limit,
 the QME is metastable with four slow eigenvalues, $m=4$: in fact, the model has a metastable DFS~\cite{Macieszczak2016}.
These four slow eigenvalues and the metastable DFS are also preserved on replacing 
\beq
H \to H + \Omega_{\rm r}\sigma_1^x\sigma_2^y
\label{equ:Omegar}
\eeq
and fixing $\Omega_{\rm r}=O(1)$ as $\epsilon\to0$.  This generates fast unitary evolution of the quantum state within the DFS.  Our analysis focuses on the model \eqref{DFS_H_J} which corresponds to $\Omega_{\rm r}=0$, but we also include brief comments on the general case.

We recall the distinction of~\cite{Macieszczak2016} between classical metastability and a metastable DFS.  In classical metastability for $m=4$, the QME solution \eqref{MS_approx} for $t\gg\tauf$ would become
\beq
\rho(t) \approx \sum_{X\in\{A,B,C,D\}} p_X(t) \rho_X
\label{equ:mani-classical-m4}
\eeq
where
the four states are $A,B,C,D$, the $p_X(t)$ are time-dependent probabilities of each state (summing to unity), and $\rho_X$ is a density matrix describing phase $X$~\cite{Macieszczak2016,classical_metastability_in_quantum}.
 All these density matrices are supported on different parts of the system's Hilbert space.  The four terms in the sum mirror the $m$ terms in \eqref{MS_approx}.
In contrast, for the model considered here, we have (at the same level of accuracy):
\begin{multline}
\rho(t) \approx p_1(t) \ket{\!\uparrow\downarrow} \bra{\uparrow\downarrow\!} + p_2(t) \ket{\!\downarrow\uparrow} \bra{\downarrow\uparrow\!}
\\ + z(t) \ket{\!\uparrow\downarrow} \bra{\downarrow\uparrow\!}  + z^*(t) \ket{\!\downarrow\uparrow}\bra{\uparrow\downarrow\!}
\label{equ:mani-dfs}
\end{multline}
where now $p_{1,2}(t)$ are real-valued probabilities (summing to unity) but $z(t)$ is complex (with $|z|^2 \leq p_1p_2$).
This means that the metastable manifold approximately corresponds to a qubit with basis states $\ket{\xup\xdown}$ and $\ket{\xdown\xup}$.
The $z$-terms in \eqref{equ:mani-dfs} represent coherences between different parts of the metastable manifold, which are forbidden in \eqref{equ:mani-classical-m4}.

\subsection{Connections between trajectory and QME dynamics}

Figs.~\ref{fig:dfs_traj+record} show a typical quantum trajectory of this model, and its corresponding time record.  Compared with the models considered so far, there are several striking features.  
First, the projection of $\psi$ onto the states $\ket{\xup\xup}$ and $\ket{\xdown\xdown}$ is extremely small throughout [in fact, $O(\epsilon)$].
Second, quantum jumps are rare, throughout the evolution.
Third, the two-phase behaviour of Figs.~\ref{fig:3_state:unravelled_trajectory} and \ref{fig:3_state_2_jump:unravelled_trajectory} is absent: instead one sees occasional fast events (triggered by jumps), which are followed by periods of continuous slow evolution towards an asymptotic (dark) state.

These aspects of the behaviour are linked to properties of the QME, which are clearest from a perturbative analysis about $\epsilon=0$.  As noted above, the QME behaviour for $\epsilon=0$ corresponds to fast relaxation into \eqref{equ:mani-dfs}.  All contributions to $\rho(t)$ that involve $\ket{\xup\xup}$ and $\ket{\xdown\xdown}$ decay quickly so they do not appear in \eqref{equ:mani-dfs}, this is a decay subspace~\cite{Baumgartner_2008,lindblad_geometry_Albert}.  Similarly, quantum trajectories for $\epsilon=0$ relax into pure states $\ket{\psi}\bra{\psi}$ with $\ket\psi = z_1\ket{\xup\xdown} + z_2\ket{\xdown\xup}$, which are stationary in the unravelled dynamics at $\epsilon=0$. The manifold~\eqref{equ:mani-dfs} is annihilated by jump operators $J_1,J_2$, hence these states are dark.

On restoring a positive $\epsilon>0$, one naturally recovers small contributions of $\ket{\xup\xup},\ket{\xdown\xdown}$ in the quantum trajectories.
For small positive $\epsilon$, the small contributions of $\ket{\xup\xup},\ket{\xdown\xdown}$ restore small finite jump rates, as observed. The slow continuous evolution of the conditional state in Fig.~\ref{fig:dfs_unravelled_trajectory} is a more subtle feature of the metastable DFS.  We explain below that it is related to the evolution of coherences $z(t)$ within the metastable manifold.  Note also that this slow evolution is a feature of this model that is not fully generic for metastable DFSs: for the model of~\eqref{equ:Omegar}, Eq.~\eqref{equ:mani-dfs} includes a fast unitary evolution.
However, we focus here on $\Omega_{\rm r}=0$, for simplicity.

\begin{figure}
    \centering
    \includegraphics[width=0.48\textwidth]{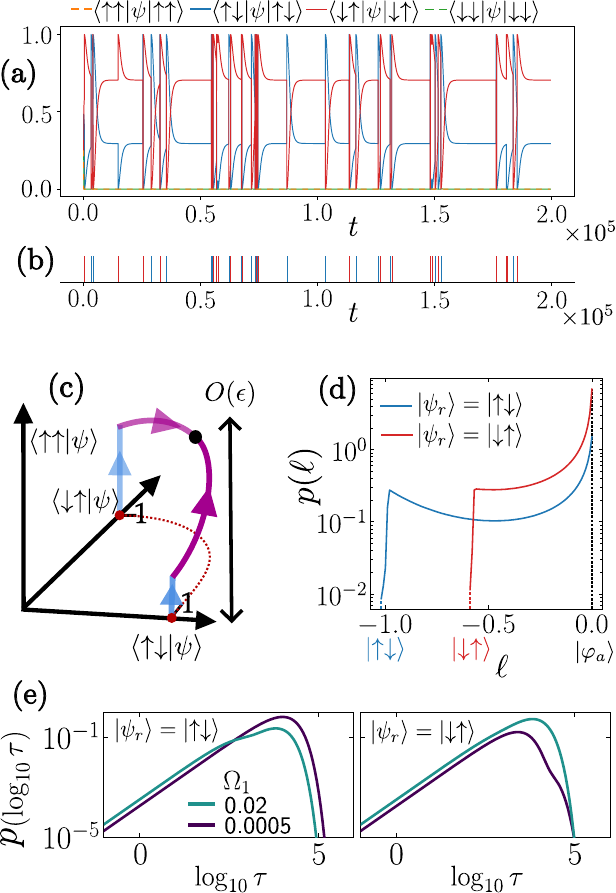}
    \caption{Two-qubit quantum reset model with metastable DFS. (a)~Example unravelled trajectory, from initial state $\frac{1}{4}\sum_{ij}|i\rangle\langle j|$. (b)~Corresponding measurement record. (c) Illustration of the unravelled dynamics. (d) Stationary state probability density along the length of the jumpless trajectory from each reset state denoted by $|\psi_r\rangle$. (e) Semi-Markov probability density. All panels take $\gamma_1=4$, $\gamma_2=1$, with $\Omega_1=0.02$, $\Omega_2=0.01$ in (a,b,d).}
    \label{fig:dfs_reset}
    \phantomlabel{(a)}{fig:dfs_unravelled_trajectory}
    \phantomlabel{(b)}{fig:dfs_measurement_record}
    \phantomlabel{(c)}{fig:dfs_jumpless_sketch}
    \phantomlabel{(d)}{fig:dfs_p(l)}
    \phantomlabel{(e)}{fig:dfs_p(log_tau)}
    \phantomlabel{(d,e)}{fig:dfs_reset_d,e}
    \phantomlabel{(a,b)}{fig:dfs_traj+record}
\end{figure}

\subsection{Quantum trajectories}\label{sec:dfs_trajectories}

To analyse 
quantum trajectories in more detail, we begin with jumpless trajectories starting from the reset points, as before.
The matrix $G$ has real elements.  For $\epsilon=0$ the states $\ket{\xup\xdown}$
and $\ket{\xdown\xup}$ are eigenvectors of $G$ with degenerate eigenvalues of zero, while $\ket{\xup\xup}$ and $\ket{\xdown\xdown}$ are eigenvectors with positive eigenvalues.  For $\epsilon>0$ the degenerate pair are split, leading to a dominant eigenvector $\ket{\varphi_a}$ and a second eigenvector $\ket{\varphi_2}$ which are both of the form $\cos\phi\ket{\xup\xdown} + \sin\phi\ket{\xdown\xup} + O(\epsilon)$, with eigenvalues $O(\epsilon^2)$.  

It follows that the jumpless trajectories evolve quickly away from the reset points before arriving in a slow manifold spanned by $\ket{\varphi_a}$ and $\ket{\varphi_2}$.  This is illustrated in Fig.~\ref{fig:dfs_jumpless_sketch}: there is fast motion along the blue lines, followed by slow motion along the purple line, which is the slow manifold.  The dotted red line can be parameterised as $\cos\vartheta\ket{\xup\xdown} + \sin\vartheta\ket{\xdown\xup}$; similarly the purple line is $\cos\vartheta\ket{\varphi_a} + \sin\vartheta\ket{\varphi_2}$: the separation of the two lines is $O(\epsilon)$.

The dominant eigenvector $\ket{\varphi_a}$ is the black dot, and both jumpless trajectories eventually converge to this point.  The two slow eigenvalues of $G$ are both $O(\epsilon^2)$, and this sets the speed at which the jumpless trajectory moves along the slow manifold. 
(Note that the jumpless trajectory describes a four-dimensional wavefunction with real co-efficients: we represent it by a three-dimensional sketch, bearing in mind that the fourth component is fixed by normalisation.)

In contrast to the models discussed so far, the properties outlined in Sec.~\ref{sec:metastable_classical_systems} do not hold in this case, and one cannot identify fast relaxation into distinct metastable phases with slow transitions between them. 
This qualitative difference from the 3-state model illustrates that the metastable DFS is inherently non-classical.  The absence of distinct phases also means that committors -- which are characteristic of classical metastability -- cannot be defined in this case.
The possibility of continuous slow evolution of the quantum state within a slow manifold is a distinctive feature of quantum metastable systems: this behaviour is absent from classical theories of metastability.

Fig. \ref{fig:dfs_p(l)} shows the invariant measure along the jumpless trajectories: there is significant probability [$p(\ell)=O(1)$ as $\epsilon\to0$] over a range of $\ell$.  
The typical time between jumps is long, as shown in Fig. \ref{fig:dfs_p(log_tau)}.  In fact, the {entire} slow manifold is similar to a dark phase in that the jump rate is $O(\epsilon^2)$ everywhere. Following the prescription of Sec.~\ref{sec:committor_for_reset}, one can use the semi-Markov property of the quantum trajectories to compute the probability that the system jumps to a given reset state given the destination of the previous jump, using \eqref{committor_from_jumps}. Since there are two reset points, there are four such probabilities: we find that they are all of order unity.  This is in stark contrast to the three-state model of Sec.~\ref{sec:3_state_2_jump} where the system tends to make multiple repeated jumps of the same type: in that case, two of the probabilities are $O(1)$ but the other two are small, $O(\epsilon^2)$.

Having described trajectories that start from the reset points, we briefly consider other initial points. The purple line in Fig.~\ref{fig:dfs_jumpless_sketch} is just one part of a larger slow manifold that corresponds to a qubit 
$\ket\psi = z_a\ket{\varphi_a} + z_2\ket{\varphi_2}$ where $z_a,z_2$ are complex in general.  A trajectory initialised anywhere on this manifold will evolve slowly towards $\ket{\varphi_a}$, and its jump rate is small [in fact, $O(\epsilon^2)$] throughout this evolution.  
The slow time scales for these processes mean that any coherence between $\ket{\xdown\xup}$ and $\ket{\xup\xdown}$ in the initial condition  survives a long time: this is a central feature of the metastable DFS.
In the specific case of quantum reset processes, jumps necessarily remove the coherences between the reset states, with the result that DFSs must always be ``dark'', as in this example.

Finally, it is also interesting to analyse quantum trajectories for the generalised model of \eqref{equ:Omegar}.  The essential difference in this case is that the purple line in Fig.~\ref{fig:dfs_jumpless_sketch} is extended to a closed circle around which the jumpless trajectories circulate quickly (frequency of order $\Omega_{\rm r}$), although the DFS remains dark, as it must.
Similarly to introducing \eqref{equ:Omegar}, one can also apply any unitary operation (gate) acting as a rotation within the slow manifold. This produces a logical qubit, from the two physical qubits, which supports long lived coherence.
The evolution under the applied operations remains approximately unitary on the long times between jump events.

\begin{figure}[t]
    \centering
    \includegraphics[width=0.45\textwidth]{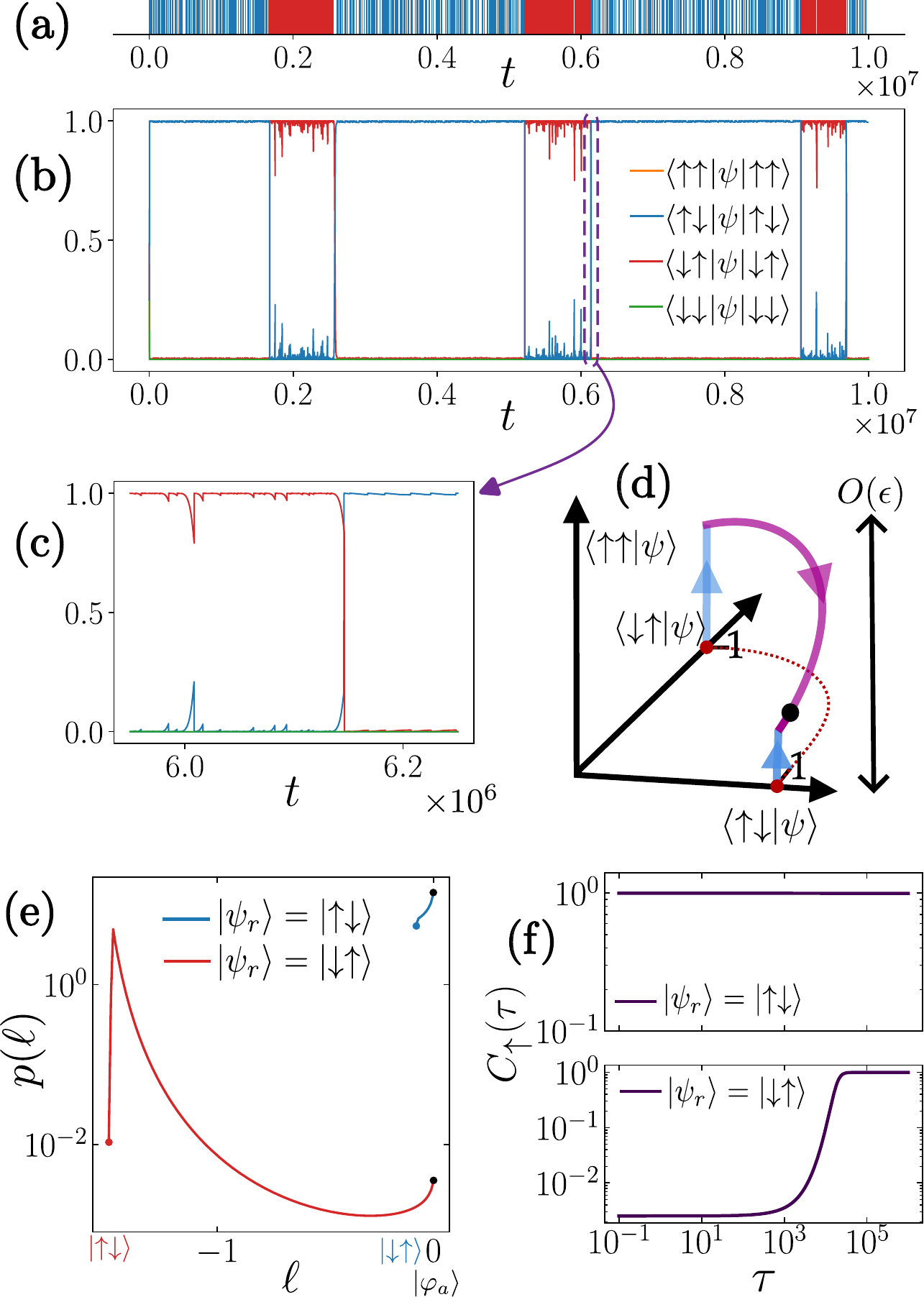}
    \caption{Two-qubit model with classical metastability. (a) Example measurement record for $\Omega_1=0.0005$, $\Omega_2=0.01$ and initial state $\frac{1}{4}\sum_{i,j}\ket{i}\bra{j}$.  (b) Corresponding quantum trajectory. (c) Magnified portion of (b). (d) Illustration of the unravelled dynamics. (e) Probability density along the length of the jumpless trajectory from each reset state denoted by $|\psi_r\rangle$. (f)~Committor to the state $\ket{\psi}\approx\ket{\xup\xdown}$ in the semi-Markov representation, calculated from \eqref{eqn:committor_reset_core_sets}.} 
    \label{fig:dfs_reset_futher_separation}
    \phantomlabel{(a)}{fig:dfs_FS_measurement_record}
    \phantomlabel{(b)}{fig:dfs_FS_example_trajectory}
    \phantomlabel{(c)}{fig:dfs_FS_zoomed_trajectory}
    \phantomlabel{(d)}{fig:dfs_FS_jumpless_sketch}
    \phantomlabel{(e)}{fig:dfs_FS_p(l)}
    \phantomlabel{(f)}{fig:dfs_FS_committor}
    \phantomlabel{(a,b)}{fig:dfs_FS_trajectory+record}
\end{figure}

\subsection{Recovering classical metastability for $\Omega_1 \ll \Omega_2$}\label{sec:dfs_recovering_classical}

In addition to a metastable DFS, the two qubit model \eqref{DFS_H_J} also supports other metastable structures.  To illustrate this, note that for $\Omega_1=0$, the Hilbert space is disconnected into two subspaces that are not mixed under the dynamics: suitable bases for the subspaces are $\{\ket{\xup\xup},\ket{\xup\xdown}\}$ and $\{\ket{\xdown\xup},\ket{\xdown\xdown}\}$.  We label these subspaces as $\uparrow$ and $\downarrow$, according to the state of the first spin. It follows that for $|\Omega_1|\ll |\Omega_2|,\gamma_1,\gamma_2$, one recovers two-phase metastability, with one phase supported on each subspace, and phenomenology similar to the 3-state model considered above.

Similarly to the analysis of Sec.~\ref{sec:relation_committor_QME}, the committors for these phases can be identified from \eqref{equ:CAP}. Labelling the phases according the subspaces, the committor for the $\uparrow$-phase can be identified as 
\begin{align}
C^*_\xup(\psi) & =  {\rm Tr}[n_1\psi]  \nonumber\\
& = \bracketmatrix{\xup\xup}{\psi}{\xup\xup} + 
\bracketmatrix{\xup\xdown}{\psi}{\xup\xdown}
\end{align}
and similarly $C^*_\downarrow(\psi)={\rm Tr}[(1-n_1)\psi] $.

A particularly interesting situation occurs for 
\beq
|\Omega_1| \ll |\Omega_2| \ll \gamma_1, \gamma_2
\eeq
in which case the QME spectrum has two gaps.  As an example, we take $\Omega_1=\epsilon^2\omega_1$, $\Omega_2=\epsilon\omega_2$ in which case the slow eigenvalues of the QME are $\lambda_1=0$, $\lambda_2=O(\epsilon^4)$, $\lambda_{3,4}=O(\epsilon^2)$.   For times $t\gg \epsilon^{-2}$, the eigenvalues $\lambda_{3,4}$ do not contribute and  one does indeed observe simple two-phase (classical) metastability.  
However, the eigenvalues $\lambda_{3,4}$ affect the behaviour on intermediate time scales.  [The system still features the ``slow manifold'' \eqref{equ:mani-dfs}. After each jump, the relaxation into this manifold occurs quickly; while motion within this manifold can occur with intermediate rates which scale as $\lambda_{3,4}$ or with slow rates, scaling as $\lambda_2$.]
This resulting situation is illustrated in Fig.~\ref{fig:dfs_reset_futher_separation}.  The two-phase metastability is clear from the representative trajectory and corresponding measurement record in Figs.~\ref{fig:dfs_FS_trajectory+record} and the existence of an intermediate relaxation time is visible in Fig.~\ref{fig:dfs_FS_zoomed_trajectory}. 

Fig.~\ref{fig:dfs_FS_jumpless_sketch} illustrates that
the stationary point $\ket{\varphi_a}$ of the unravelled dynamics is close to the reset point $\ket{\xup\xdown}$, in contrast to the situation shown in Fig.~\ref{fig:dfs_jumpless_sketch}. 
The $\uparrow$-phase consists of rapid motion from $\ket{\xup\xdown}$ towards $\ket{\varphi_a}$, with frequent jumps back to $\ket{\xup\xdown}$, reminiscent of the bright phase in the three-state models of Sec.~\ref{sec:three_state_models}.  Trajectories in the $\downarrow$-phase evolve quickly from the reset state $\ket{\xdown\xup}$ to the slow manifold: they move towards $\ket{\varphi_a}$ with an intermediate rate, but they typically reset back to $\ket{\xdown\xup}$ before getting close to $\ket{\varphi_a}$. 
These two phases are visible as two separate peaks in the distribution $p(\ell)$ in Fig.~\ref{fig:dfs_FS_p(l)}.

Finally, Fig.~\ref{fig:dfs_FS_committor} shows the committor as a function of the time since the last reset.  The core sets for the two phases are the reset points, as in Sec.~\ref{sec:3_state_2_jump}.  {Hence the committor to the $\uparrow$-phase is given by \eqref{eqn:committor_reset_core_sets} on replacing $A\to\,\uparrow$, $\kappa_A\to\gamma_1$ and $\ket{\xi_A}\to\ket{\!\uparrow\downarrow}$.} Fig.~\ref{fig:dfs_FS_committor} shows that if the last reset was to $\ket{\xup\xdown}$, the next reset state will almost certainly be the same, independent of the time since the last reset.  If the last reset was to $\ket{\xdown\xup}$, the most likely destination of the next jump depends on the time $\tau$.  For small $\tau$ then the system most likely resets to the same destination, but for large $\tau$ the state approaches $\ket{\varphi_a}$ and tends to jump to $\ket{\xup\xdown}$ instead.  This behaviour is similar to Fig.~\ref{fig:3_state_2_jump:SM_committor}.

Since this model has frequent jumps in both phases, it is essential for the two-phase metastability that the destination of each jump tends to be similar to that of the previous one (recall Sec.~\ref{sec:3_state_2_jump}).
To see how this happens, it is convenient to write the slow manifold as
$\ket{\psi}=\cos\vartheta\ket{u} + \sin\vartheta\ket{v}$ with $\ket u=\ket{\xup\xdown} +O(\epsilon)$, and $\ket v=\ket{\xdown\xup} +O(\epsilon)$.  The $\uparrow$-phase corresponds to $\vartheta\approx 0$ and it can be shown that $\bracket{\xup\xup}{u}=O(\epsilon)$, while $\bracket{\xdown\xdown}{u}=O(\epsilon^2)$.
Constructing the semi-Markov rates from~\eqref{sm_jump_rate}, one finds that systems in the $\uparrow$-phase jump to $\ket{\xup\xdown}$ with rate $O(\epsilon^2)$ but they jump to $\ket{\xdown\xup}$ with a much smaller rate $O(\epsilon^4)$. Hence, resets to the same state are indeed much more frequent.  A similar situation holds in the $\downarrow$-phase.

\section{OUTLOOK}\label{sec:outlook}

\subsection{Beyond quantum reset models}\label{sec:non_reset_dfs_model}

All examples considered so far have been quantum reset models: we emphasised that quantum trajectories are particularly tractable in this case.  However, the metastable phenomenology discussed above is not restricted to these models.  In particular, the relationship \eqref{equ:CAP} between the committor and the spectrum of the QME is generic for system with classical metastability (in the sense of~\cite{classical_metastability_in_quantum}).

We offer two specific examples of metastable non-reset processes.  The first is obtained by 
combining the jump operators of the two-jump three-state model of Sec.~\ref{sec:3_state_2_jump} into a single jump operator
\beq
    J = J_1+J_2 = \sqrt{\kappa_1}|0\rangle\langle 1| + \sqrt{\kappa_2}|2\rangle\langle 2|.
\eeq
This operator does not mix the unperturbed phases, so the coupling between them is unchanged from the original model and the low-lying spectrum of the QME has the same qualitative features.  As a result, the committors to the two phases are still given to leading order by \eqref{equ:CB-CD-formulae}, and the average jump rates within the phases are also unchanged, although the higher-order statistics are different.  The quantum trajectories have the same qualitative behaviour as the original quantum reset model. In this sense, the reset model is a tractable system that exemplifies a broader class of metastable systems.

A similar construction can be used to analyse a two-qubit model
introduced in~\cite{Macieszczak2016}.  This has the same Hamiltonian as~\eqref{DFS_H_J}, but a single jump operator
\beq\label{supo_jump_op}
J = \sqrt{\gamma_1}n_1\sigma_2^--\sqrt{\gamma_2}(1-n_1)\sigma_2^+
\eeq 
which replaces the two jumps in~\eqref{DFS_H_J} by a single jump which superposes them.
The low lying spectrum of the QME is almost unchanged since at $\epsilon=0$ the DFS is dark, so that at $\epsilon>0$ the QME spectrum has $m=4$ with the metastable DFS.

\begin{figure}
    \centering
    \includegraphics[width=0.48\textwidth]{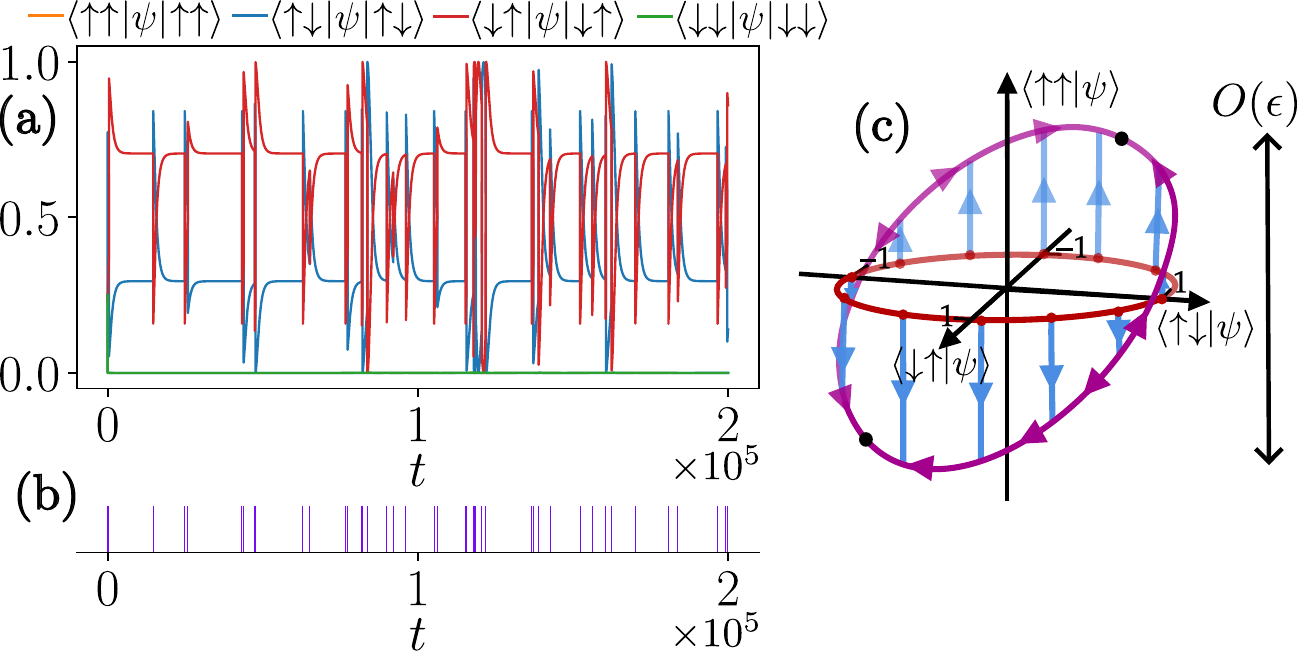}
    \caption{Non-quantum-reset variant of two-qubit model. (a) Example unravelled trajectory, from initial state $\frac{1}{4}\sum_{i,j}|i\rangle\langle j|$ with $\Omega_1=0.02$, $\Omega_2=0.01$, $\gamma_1=4$, $\gamma_2=1$.  (b) Measurement record for (a). (c) Illustration of the dynamics (see text).
    }
    \label{fig:dfs_non_reset}
    \phantomlabel{(a)}{fig:dfs_non_reset_qme}
    \phantomlabel{(b)}{fig:dfs_non_reset_unravelled}
    \phantomlabel{(c)}{fig:dfs_non_reset_graphic}
\end{figure}

Fig.~\ref{fig:dfs_non_reset} shows a representative quantum trajectory for this model, with its corresponding measurement record, and a sketch illustrating the behaviour of jumpless trajectories.  This can be compared with Fig.~\ref{fig:dfs_reset}.  The main difference is that the system is not a quantum reset model:  the red circle in Fig.~\ref{fig:dfs_non_reset_graphic} shows the possible destinations of quantum jumps, which are all states of the form $\ket{\psi}=q\ket{\xup\xdown} \pm \sqrt{1-q^2} \ket{\xdown\xup}$ with $-1\leq q\leq 1$.\footnote{%
Since the state of the unravelled dynamics is the pure density matrix $\psi=\ket\psi\bra\psi$, the points $\ket\psi$ and $-\ket\psi$ are equivalent in this graphical illustration.} 

From \eqref{supo_jump_op}, the jump destination (parameterised by $q$) depends on the small components of $|\psi\rangle$, which have a non-trivial evolution, leading to a complicated dependence of $q$ on the starting point of the jump.
However, it is still useful to consider jumpless trajectories in this representation.  Every jump takes the system onto the red circle, after which the state relaxes quickly onto the slow manifold (purple circle).  It then evolves within that manifold towards the asymptotic state $\ket{\varphi_a}$.  Jumps may take place from any point on the slow manifold and may end at any point on the red circle.  As a result, there is a continuous family of jumpless trajectories.  In contrast to the corresponding quantum reset model (Section \ref{sec:two_qubit_model}), this means that the steady state probability distribution cannot be represented as in Fig.~\ref{fig:dfs_reset_d,e}.  However, the DFS phenomenology is very similar between reset and non-reset models: Again, the reset model is useful as a tractable example that illustrates generic metastable behaviour.  
An interesting question is whether there are qualitative aspects of metastability that cannot be captured by reset models: an important case might be the metastability that appears near first-order phase transitions in many-body systems~\cite{Minganti2018,Landa20,meta_ising}.

\subsection{Quantum Systems with Classical Metastability}

A central conclusion of Secs.~\ref{sec:three_state_models} and \ref{sec:metastable_unravelled} is a connection between quantum trajectories and those of classical systems with metastability.  This appears in  quantum systems with classical metastability in the sense of \cite{Macieszczak2016,classical_metastability_in_quantum}.  The associated phenomenology is that  typical trajectories relax quickly into one of the metastable phases and explore them quickly; on long time scales they exhibit rare transitions between the phases.  In the stationary regime, these trajectories show intermittent behaviour, which switches between the phases: this mirrors the intermittent behaviour of time records discussed in~\cite{classical_metastability_in_quantum}.
The rates for the slow and fast processes in such systems are controlled by $\taumet$ and $\tauf$ respectively.
In the classical framework, an important role is played by the committor: for two-phase metastability, we explained that this can be related to the spectrum of the QME, via~\eqref{equ:CAP}, providing quantitative connection between QME and trajectory representations.

As examples of this behaviour we considered two variants of a simple 3-state model, which exhibits bright and dark states.  Both examples are quantum reset processes: we explained how this structure can be exploited in order to characterise their behaviour via a mapping to semi-Markov processes, and by analysis of jumpless trajectories that start from the reset point(s).  Hence the stationary state of the unravelled dynamics is supported on a finite set of lines in the Hilbert space. 
Our results show these quantum reset models are relatively simple to analyse, while still supporting rich metastable behaviour.  This means that some general insights can be extrapolated from such simple examples: for example, metastable systems generically show large differences between quantum trajectories and the averaged behaviour of the QME; we also shows how the committor is useful for delineating metastable phases within quantum trajectories.


To continue the theoretical programme that we have started here, it would be important to characterise more rigorously the conditions required to observe intermittency in unravelled dynamics.
One could also investigate models which are metastable but do not exhibit this intermittent behaviour, for example a qubit with $H=\Omega\sigma^z$, ${J = \epsilon\gamma_x\sigma^x + \epsilon\gamma_y\sigma^y + \gamma_z\sigma^z}$.
Other interesting questions occur in quantum trajectories where two metastable phases have the same average jump rate, which makes them harder to distinguish from experimental time records. 

\subsection{Quantum Systems with Non-Classical Metastability}
Our analysis of the two-qubit model of Sec.~\ref{sec:two_qubit_model} illustrates the observation of~\cite{Macieszczak2016}, that quantum systems can support non-classical metastable behaviour such as metastable DFS.
%
  Consistent with the nomenclature of non-classical metastability, we find that the corresponding quantum trajectories do not exhibit the classical metastable phenomenology of fast relaxation into metastable phases and rare transitions between them.  Instead, the unravelled dynamics features a slow manifold within which they undergo continuous (slow) motion.

The existence of metastable DFS in quantum reset models may be surprising, given the simplicity of the models.  We explained that this is possible because the DFS is ``dark'', in that all jump rates vanish as $\epsilon\to0$, within the relevant subspace.  This is also true for the non-reset model discussed in Sec.~\ref{sec:non_reset_dfs_model}.

Quantum trajectories are highly relevant to experimental setups, especially where systems undergo continuous monitoring as is becoming increasingly implemented in modern experiments \cite{Geremia2004,Katz2006,Campagne_Ibarcq2016}. Therefore, understanding their behaviour in non-classically metastable systems, which exhibit phenomena crucial for implementation of quantum technologies, is an important task. We have investigated this in few-level systems, both reset and non-reset, where the trajectories behave analogously to slow continuous classical dynamics within a manifold of low activity. A natural next step is to build on this work to consider many-body systems, which have a greater range of practical applications \cite{Kempe2001,Ahn2003,Pastawski2011}.

\begin{acknowledgments}
    We thank Juan P. Garrahan for helpful discussions about quantum trajectories and metastable systems.
\end{acknowledgments}

\appendix

\section{Three-state model}
\label{appendix:3_state}

\subsection{{Quantum Master Equation}}
\label{app:lind-3state}

{Here we discuss the structure of the QME for the model of Sec.~\ref{sec:three_state_intro}. Writing the QME operator in the basis
$$\{|2\rangle\langle2|, |1\rangle\langle1|, |0\rangle\langle0|,\sigma_{01}^x,\sigma_{02}^x,\sigma_{12}^x,i\sigma_{01}^y,i\sigma_{02}^y,i\sigma_{12}^y\},$$
where $\sigma_{jk}^x=|j\rangle\langle k|+|k\rangle\langle j|$ and $i\sigma_{jk}^y=|j\rangle\langle k|-|k\rangle\langle j|$, we obtain a matrix representation of ${\cal L}$ as
}

\definecolor{brightblue}{RGB}{31, 119, 180}
\newcommand{\bb}[1]{\color{brightblue}#1}
\definecolor{darkgreen}{RGB}{31, 119, 180} 
\newcommand{\dg}[1]{\color{darkgreen}#1}
\small
\beq\label{QME_matrix} 
    {{L}} = \begin{pNiceMatrix}[margin]
    \Block[fill=red!25,rounded-corners]{6-6}{}0&0&0&0&{\color{black}-\Omega_2}&0&0&0&0\\
    0&-\kappa_1&0&-\Omega_1&0&0&0&0&0\\
    0&\kappa_1&0&\Omega_1&\Omega_2&0&0&0&0\\
    0&2\Omega_1&-2\Omega_1&-\frac{\kappa_1}{2}&0&\Omega_2&0&0&0\\
    \color{black}2\Omega_2&0&{-2\Omega_2}&0&0&\Omega_1&0&0&0\\
    0&0&0&-\Omega_2&-\Omega_1&-\frac{\kappa_1}{2}&0&0&0 \RowStyle[cell-space-bottom-limit=1.5mm]{}\\
    0&0&0&0&0&0& \Block[fill=blue!25,rounded-corners]{3-3}{} -\frac{\kappa_1}{2}&0&-\Omega_2\\
    0&0&0&0&0&0&0&0&\Omega_1\\
    0&0&0&0&0&0&\Omega_2&-\Omega_1&-\frac{\kappa_1}{2} \RowStyle[cell-space-bottom-limit=0.3mm]{}
    \end{pNiceMatrix}
\eeq
\normalsize
This matrix has a block structure, the steady state eigenvector comes from the first block while the eigenvalues of the second block have strictly negative real parts and correspond to decaying coherences.

To understand metastability at this level, recall that $\Omega_2=\epsilon\omega_2$ and consider perturbation theory about $\epsilon=0$ by writing ${{L}}=L_0+\epsilon V$. The first row and column of $L_0$ are both full of zeros, showing that $\rho_{D}^*=\ket{2}\bra{2}$ is a (dark) steady state of the model with $\epsilon=0$.  There is a second (bright) steady state $\rho_{B}^*$ which is supported on $\ket{1}\bra{1}$, $\ket{0}\bra{0}$, and $\sigma^x_{01}$.

For small positive $\epsilon$ one recovers a unique stationary state which is well-approximated by a linear combination of $\rho_{B}^*$ and $\rho_{D}^*$.
The two zero eigenvalues are split as $\theta_1=0$ and $\theta_2=O(\epsilon^2)$.
(One might expect in general $\theta_2=O(\epsilon)$ but there is no such contribution here because $\Tr[\rho_{B}^*V(\rho_{D}^*)]=0$.)

\subsection{Jumpless trajectories}
\label{app:G-3state}

We describe here the jumpless dynamics of the three-state model, as defined in~\eqref{equ:cond3}.  For this, we require some properties of $G$ (or equivalently the effective Hamiltonian $H_{\rm eff}=iG$).
Working in the basis $(\ket{0},\ket{1},\ket{2})$, we have
\beq
    G=\begin{pmatrix}
        0&\Omega_1&\Omega_2\\
        -\Omega_1&-\frac{\kappa_1}{2}&0\\
        -\Omega_2&0&0
    \end{pmatrix}.
\eeq
Using also that
$\Omega_2=\epsilon\omega_2$, the system can be analysed perturbatively in $\epsilon$.  
For $\epsilon=0$, the only non-zero elements of $G$ appear in a $2\times2$ block: the corresponding subspace of the model's Hilbert space corresponds to the bright phase and the effective Hamiltonian is that of a two-level atom.  The eigenvectors of $G$ at $\epsilon=0$ are $ |\varphi_a^*\rangle = |2\rangle$ (the dark state) with eigenvalue $\theta_a^*=0$, and 
\beq\label{2_lvl_eigvecs}
|\varphi_\pm^*\rangle \propto{ \theta^*_\mp|0\rangle+\Omega_1|1\rangle } \; ,
\eeq 
where the coefficient of proportionality is fixed by normalisation and 
\beq\label{2_lvl_eigvals}
\theta_\pm^* = \frac{-\kappa_1\pm\sqrt{\kappa_1^2-16\Omega_1^2}}{4}
\eeq
are the corresponding eigenvalues, which are of order unity.
For $\epsilon>0$, the 
eigenvectors of $G$ are 
\begin{align}
|\varphi_\pm\rangle & =|\varphi_\pm^*\rangle+O(\epsilon)
\nonumber\\
|\varphi_a\rangle & = |2\rangle+O(\epsilon).
\label{equ:eigG}
\end{align} 
The corresponding eigenvalues are 
$\theta_\pm = \theta^*_\pm + O(\epsilon)$ and $\theta_a=O(\epsilon^2)$.
The perturbative corrections in~\eqref{equ:eigG} are responsible for weak coherences in the EMSs, as described in Sec.~\ref{sec:connection_trajectories_QME}.

The eigenvalues $\theta_\pm$ may be real, or a complex conjugate pair.  We focus on the case $\kappa\geq4\Omega_1$ in which case they are both real.  [In our numerical examples we take $\kappa=4\Omega_1$ in which case $\theta_\pm^*$ are degenerate, but they are split by the perturbation, so $\theta_\pm$ are real and distinct.  Hence, the numerics reflect qualitatively what occurs for all $\kappa_1\geq 4\Omega_1$.]

Restricting to this case we have $0>\theta_a>\theta_+>\theta_-$ and \eqref{equ:cond3} becomes
\begin{equation}
    \ket{\psi_\tau} \propto A_a 
    \ket{\varphi_a}+A_+\ket{\varphi_+}e^{-(\theta_a-\theta_+)\tau}+A_-\ket{\varphi_-}e^{-(\theta_a -\theta_-)\tau}
   \label{equ:jumpless3-spectrum}
\end{equation}
for suitable (real) constants $A_a,A_\pm$; 
the constant of proportionality is set by normalisation and we assume $A_a>0$ without loss of generality.
Since the jumpless trajectory starts from the reset state $\ket{0}$, we must have $A_a|\varphi_a\rangle+A_+|\varphi_+\rangle+A_- |\varphi_- \rangle=|0\rangle$.  Note that
$G$ is not Hermitian, so its eigenvectors are not orthogonal.  Nevertheless, multiplying from the left by $\bra{2}$ and using \eqref{equ:eigG} shows that $A_a=O(\epsilon)$, and similarly the overlaps with $\bra{0}$ and $\bra{1}$ show that  $A_\pm=O(1)$.

The exponential factors in \eqref{equ:jumpless3-spectrum} are decaying with rates of $O(1)$, so for very long times we must have $\ket{\psi_\tau} \approx \ket{\varphi_a}$.  However, the small coefficient $A_a=O(\epsilon)$ means that this only occurs for times $\tau\gtrsim (\theta_+-\theta_a)^{-1}\log(|A_+|/\epsilon)$ which diverges as $\epsilon\to0$.  On the other hand, $(\theta_+-\theta_-)=O(1)$ so one sees that for $(\theta_+-\theta_-)^{-1}\ll \tau \ll (\theta_+-\theta_a)^{-1}\log(|A_+|/\epsilon)$  one has $\ket{\psi_\tau} \approx \ket{\varphi_+}$.
This state corresponds to the elbow in Fig.~\ref{fig:3_state_panel2}: the jumpless trajectory moves slowly past this elbow because the contribution of $\ket{\varphi_-}$ decays almost to zero before the contribution of $\ket{\varphi_a}$ becomes significant.  

\newcommand{\dmac}{d}

It is useful to identify the time $\tau_{\rm e}$ at which the jumpless trajectory ``passes the elbow''.  We define this via the relation
\beq
A_a  = \dmac |A_+| e^{-(\theta_a-\theta_+)\tau_{\rm e}} 
\label{equ:a-thresh}
\eeq
where $\dmac>0$ is a threshold that describes how far past the elbow the trajectory must go: for times $\tau>\tau_{\rm e}$, the relative contribution of $\ket{\varphi_a}$ to $\ket{\psi_\tau}$ is at least $\dmac$.  Recalling that $A_a=O(\epsilon)$ then $\tau_{\rm e}$ diverges as $\log(1/\epsilon)$.  Now observe that 
once $\ket{\psi_\tau}$ passes the elbow, the jumpless trajectory converges exponentially fast into the dark state $\ket{\varphi_a}$ with rate $(\theta_+-\theta_a)^{-1}=O(1)$.  This means that while transitions from the elbow to the dark phase are rare (see below), the actual transition takes place quickly.

To understand why these transitions are rare, it is useful to estimate the probability that a trajectory starting from the reset state $\ket{0}$ does actually pass the elbow before jumping back to $\ket{0}$.  This is exactly the survival probability of \eqref{survival_probability}, evaluated at $\tau_{\rm e}$.  From the properties of $G$, this results in
\beq
S(\tau_{\rm e}) \approx A_a^2 e^{2\theta_a \tau_{\rm e} }+ A_+^2 e^{2\theta_+\tau_{\rm e}},
\eeq
where the approximate equality is due to subleading corrections from (fast-decaying) exponential factors $e^{-(\theta_a -\theta_-)\tau}$ and the fact that $\bracket{\varphi_a}{\varphi_+}=O(\epsilon)$.  Using \eqref{equ:a-thresh} we obtain
$
S(\tau_{\rm e}) \approx A_a^2 (1+\dmac^{-2}) {\rm e}^{2\theta_a\tau_{\rm e}} 
$. 
For small $\epsilon$ then $A_a=O(\epsilon)$ and $\dmac=O(1)$ and $\theta_a\tau_{\rm e}\sim \epsilon^2\log(1/\epsilon)\to0$ so the probability that the conditional state passes the elbow before jumping is 
\beq
S(\tau_{\rm e})=O(\epsilon^2).
\eeq
In other words, there are typically $O(\epsilon^{-2})$ jumps within the bright state before any transition to the dark state. 
This is consistent with the bright state being metastable.

As a final comment in this Appendix, recall that we have analysed the behaviour for $\kappa_1\geq 4\Omega_1$.  For the opposite case $\kappa_1< 4\Omega_1$, the jumpless trajectory does not feature an elbow.  For $\epsilon=0$ this trajectory describes circles in the plane $\langle 2|\psi\rangle=0$.  For $\epsilon>0$ these circles slowly spiral into $\ket{\psi_a}$.  However, the time-dependent overlap between $\ket{\varphi_a}$ to $\ket{\psi_t}$ is similar to the case already considered: it only becomes significant after a time of order $\log(1/\epsilon)$, after which the jumpless trajectory converges exponentially into $\ket{\varphi_a}$ with a rate of $O(1)$.

\bibliography{refs.bib}

\end{document}